\documentclass{emulateapj}
\usepackage{graphicx}
\def \hb{H$\beta$}
\def \haf{H$\alpha$F}
\def \ha{H$\alpha$}
\def \oiiiab{[O\,III\,$\lambda\lambda4959,5007$\,\AA]}
\def \oiiib{[O\,III\,$\lambda5007$\,\AA]}
\def \oiiia{[O\,III\,$\lambda4959$\,\AA]}
\def \oiii{[O\,III]}
\def \afe{[$\alpha$/Fe]}
\def \zh{[Z/H]}

\def \z{\phantom{0}} 
\def\lesssim{\mathrel{\hbox{\rlap{\hbox{\lower4pt\hbox{$\sim$}}}\hbox{$<$}}}}
\def\gtrsim{\mathrel{\hbox{\rlap{\hbox{\lower4pt\hbox{$\sim$}}}\hbox{$>$}}}}
\let\la=\lesssim

\newcommand{\hdf}{H$\delta_{\rm F}$}
\newcommand{\hgf}{H$\gamma_{\rm F}$}
\newcommand{\mm}[1]{\mbox{$#1$}}
\newcommand{\unit}[1]{\ifmmode \:\mbox{\rm #1}\else \mbox{#1}\fi}

\newcommand{\dage}{age $\propto \sigma$}
\newcommand{\dzhd}{$\Delta$[Z/H]/$\Delta$log$\sigma$}
\newcommand{\dzh}{Z/H $\propto \sigma$}

\newcommand{\dafe}{$\alpha$/Fe $\propto \sigma$}

\newcommand{\mone}{\mm{^{-1}}}

\newcommand{\kms}{\unit{km~s\mone}}

\newcommand{\ephot}{\mm{\epsilon_{\rm phot}}}
\newcommand{\hkpc}{\mm{h}\mone kpc}
\newcommand{\efib}{\mm{\epsilon_{\rm fib}}}
\newcommand{\eran}{\mm{\epsilon_{\rm ran}}}
\newcommand{\half}{\frac{1}{2}}
\begin{document}

\title{NOAO Fundamental Plane Survey --- II. Age and Metallicity along the Red Sequence from Linestrength Data}
\author{Jenica E. Nelan}
\affil{Department of Physics and Astronomy, Dartmouth College, Hanover, NH, USA}
\author{Russell J. Smith, Michael J. Hudson}
\affil{Department of Physics, University of Waterloo, Waterloo, Ontario, Canada \ N2L 3G1}
\author{Gary A. Wegner}
\affil{Department of Physics and Astronomy, Dartmouth College, Hanover, NH, USA}
\author{John R. Lucey, Stephen A. W. Moore, Stephen J. Quinney}
\affil{Department of Physics, University of Durham, South Road, Durham DH1 3LE, United Kingdom}
\author{Nicholas B. Suntzeff}
\affil{Cerro Tololo Inter-American Observatory, Casilla 603, La Serena, Chile}

%\affil{1}{Department of Physics and Astronomy, Dartmouth College, Hanover, NH, USA}
%\affil{2}{Department of Physics, University of Waterloo, Waterloo, Ontario, Canada \ N2L 3G1}
%\affil{4}{Cerro Tololo Inter-American Observatory, La Serena, Chile}
%\affil{3}{Department of Physics, University of Durham, South Road, Durham, United Kingdom}
%\affil{5}{Dominion Astrophysical Observatory, Victoria BC, Canada}

\slugcomment{Accepted by the Astrophysical Journal}

\begin{abstract}
We present spectroscopic linestrength data for 4097 red-sequence
galaxies in 93 low-redshift galaxy clusters, and use these to
investigate variations in average stellar populations as a function of
galaxy mass.  Our analysis includes an improved treatment of nebular
emission contamination, which affects $\sim{}10\%$ of the sample
galaxies.  Using the stellar population models of D. Thomas and
collaborators, we simultaneously fit twelve observed
linestrength$-\sigma$ relations in terms of common underlying trends
of age, [Z/H] (total metallicity) and \afe\ ($\alpha$-element
enhancement).  We find that the observed linestrength-$\sigma$
relations can be explained only if higher-mass red-sequence galaxies
are, on average, older, more metal rich, and more $\alpha$-enhanced
than lower-mass galaxies.  Quantitatively, the scaling relations are
\dage$^{0.59\pm0.13}$, \dzh$^{0.53\pm0.08}$ and \dafe$^{0.31\pm0.06}$,
where the errors reflect the range obtained using different subsets of
indices.  Our conclusions are not strongly dependent on which Balmer
lines are used as age indicators.  The derived age$-\sigma$ relation
is such that if the largest ($\sigma\sim400$\,\kms) galaxies formed
their stars $\sim13$\,Gyr ago, then the mean age of low-mass
($\sigma\sim50$\,\kms) objects is only $\sim$4\,Gyr.  The data also
suggest a large spread in age at the low-mass end of the red sequence,
with 68\% of the galaxies having ages between 2 and 8 Gyr.  We
conclude that although the stars in giant red galaxies in clusters
formed early, most of the galaxies at the faint end joined the red
sequence only at recent epochs.  
This ``down-sizing'' trend is in good qualitative agreement with
observations of the red sequence at higher redshifts, but is not
predicted by semi-analytic models of galaxy formation.
\end{abstract}

\section{Introduction}
\label{sec:intro}

Although early-type galaxies contain the bulk of the stellar mass in
the low redshift Universe, their formation histories remain poorly
understood.  Early-type galaxies lie on a tight ``red sequence'' in
the color-magnitude diagram \citep{Sandage78,Bower92} and follow
well-known dynamical scaling relations, such as the Faber-Jackson
\citep{faber76}, the $D_n - \sigma$ \citep{Dressler87}, and the
Fundamental Plane (FP) relations \citep{Djorgovski87}.  In addition to
these scaling relations, elliptical galaxies also exhibit systematic
correlations between spectroscopic absorption linestrengths and
velocity dispersion, $\sigma$. Especially well studied are the
linestrength centered on the magnesium triplet of absorption lines
near $\lambda$5175 with $\sigma$
\citep{Bender93,Wegner99,Kunt01,Bernardi03,7S}, as well as iron lines
and the Balmer series (e.g., Mehlert et al. 2003).

The tightness of these scaling relations have generally been
interpreted as evidence for coeval formation of early-type galaxies
(Bower et al. 1992), with the slope of the color-magnitude relation
arising from a mass-metallicity sequence.  However, attempts to fit
their spectra with stellar population models suggest that the
formation of early-types may be more complicated, and that there may
be a spread in their ages
\citep{Worthey94,TMB03,Trager00b,Caldwell03}.  Recent studies of
early-type galaxies at high redshift indicate that the scatter and
zero-point of the FP relation varies with redshift \citep{Wuyts04},
and may indicate that stellar population effects and age both play
roles in determining the FP \citep{vanDokkum03}.

A primary goal of the NOAO Fundamental Plane Survey (NFPS) is to
address these questions by using early-type galaxies in nearby
clusters ($z < 0.07$). The NFPS dataset consists of photometry and
spectroscopy for 5479 galaxies belonging to 93 clusters, and is ideal
for studying the evolution and properties of early-type galaxies in
the cluster environment.  The galaxies that have been selected for
spectroscopic analysis are all cluster ``red-sequence'' galaxies.
Therefore they are mostly elliptical or S0 galaxies, but there is no
explicit selection by morphology.

The NFPS data are to be presented in a series of papers. Paper I
(Smith et al. 2004; hereafter NFPS-I) described the goals and
selection of the survey and contains redshifts and velocity
dispersions.  The purpose of this paper, the second in the series, is
to correlate the linestrengths extracted from our high quality
spectroscopic data with our velocity dispersions and use the resulting
relations to investigate the broad trends of stellar age and
metallicity along the mass sequence.  Section~\ref{sec:absix} presents
the measurements of the absorption linestrengths and their errors and
in Section~\ref{sec:emiss} we detail our new method of measuring
emission lines in the \hb\ linestrength.  In Section~\ref{sec:sample}
we describe how we chose which galaxies to include in our final
sample.  Section~\ref{sec:ixsig} shows the derived
linestrength-$\sigma$ relations for many of the linestrengths, and in
Section~\ref{sec:trends}, we use stellar population models to derive
global age, metallicity and $\alpha$-element enhancement, \afe, trends
as a function of velocity dispersion.  We compare our trends to those
from other studies and to results from intermediate-redshift
observations in Section~\ref{sec:discussion}.

\section{Absorption-line Measurements}
\label{sec:absix}

\subsection{Spectroscopic Data}

NFPS-I \citep{Smith04} contains detailed descriptions of the
selection of the data for spectroscopic follow-up along with the
observations and measurements of redshifts and velocity dispersions.
To summarize, NFPS clusters were X-ray selected from the XBACS
\citep{Ebeling96} and BCS \citep{Ebeling98} catalogs for imaging.
Individual galaxies were selected for follow-up spectroscopic
observing based on their position in the cluster color-magnitude
diagram relative to the red-sequence ridge-line.  Specifically,
galaxies with $R <$ 17 and $\Delta(B-R) >$ 0.2 were chosen for
spectroscopic observation.  As noted above, there are no explicit
morphological selection criteria.

NFPS spectroscopic observations were carried out using the Hydra
multi-fiber spectrographs at the CTIO Blanco 4m telescope and the 3.5m
WIYN telescope at Kitt Peak.  Approximately 50-70 galaxies were
observed in each cluster.  Spectral resolution for both CTIO and WIYN
is {3 \AA} with CTIO data sampled at {1.15 \AA} pixel$^{-1}$ and WIYN
data sampled at {1.4 \AA} pixel$^{-1}$.  The median $S/N$ for both
WIYN and CTIO data is 22.  Variance-weighted extraction of the
spectra, along with cosmic ray rejection, wavelength calibration and
sky subtraction were performed using tasks in the HYDRA package in
IRAF.

\subsection{Linestrength measurements}
\label{subsec:rawix}

The linestrength indices used in this paper are from the original
Lick/IDS system \citep{Burstein84} and its extensions to high-order
Balmer lines \citep{Worthey97}, including H$\beta^+$
\citep{Gonzalez93}.  Note that the H$\delta_{\rm A}$ and H$\delta_{\rm
F}$ linestrength from \cite{Worthey97} differ slightly from the
H$\delta$ linestrength used in, e.g., \cite{Balogh99}; the
H$\delta_{\rm A}$ central bandpass is almost identical to their
H$\delta$, but the red and blue continua differ by $\sim$10\AA\ on
each side.  H$\delta_{\rm F}$ is a much narrower linestrength than
that of \cite{Balogh99}.

Additionally, we defined and measured two new linestrengths around the
H$\alpha$ line, which falls within the range of the WIYN spectra, for
$\sim$700 galaxies with $cz\la13000$\,\kms.  The two linestrengths,
summarized in Table~\ref{tab:halpha}, differ in their sensitivity to
contamination from neighboring [NII] emission lines. H$\alpha_{\rm F}$
has a narrow definition, with a narrow red continuum bandpass between
the [NII] 6583\,\AA\ emission line and H$\alpha$. This is primarily
designed to detect H$\alpha$ emission where present. H$\alpha_{\rm A}$
has a wider red continuum, which is contaminated by NII emission when
present. In the absence of emission, however, H$\alpha_{\rm A}$ should
be the more reliable indicator of stellar H$\alpha$ absorption.

\begin{deluxetable}{lcc}
\tablecaption{NFPS H$\alpha$ Line Indices}
\tablewidth{0pt}
\tablehead{
\colhead{Line Index}  &
\colhead{Bandpass (\AA)}    &
\colhead{Pseudocontinua (\AA)}}
\startdata
H$\alpha$F & 6554.000-6568.000 & 6515.000-6540.000 \\
	   &                   & 6568.000-6575.000 \\
H$\alpha$A & 6554.000-6575.000 & 6515.000-6540.000 \\
	   &		       & 6575.000-6585.000 \\
\enddata
\label{tab:halpha}
\end{deluxetable}

The molecular TiO$_{1}$ and TiO$_{2}$ linestrengths are only measured
in $\sim$1000 of our galaxies, most of which are from our northern
clusters since the wavelength range of the WIYN spectrograph extends
further into the red than that at CTIO.  Similarly, relatively few
galaxies have H$\alpha$ measurements for the same reason.

Prior to measuring linestrengths, the galaxy spectra were corrected to
an approximate relative flux scale by comparison with a model
elliptical template from \cite{Kinney96}.  However, because we did not
observe Lick calibration stars, we have not attempted to correct our
galaxy spectra in order to directly match to the Lick flux system.
Because each linestrength is defined using a pair of pseudocontinua
bracketing the line-feature, the linestrength indices are robust
against differences in the Lick and NFPS response functions unless
there is substantial relative curvature over the extent of a
linestrength.  The most susceptible definitions are those with the
widest extent, viz. the ``molecular'' linestrengths Mg$_{1,2}$,
TiO$_{1,2}$, and CN$_{1,2}$.  Caution should be used when analyzing
these linestrengths, especially if the sample covers a substantial
range in redshift.  Note also that the NFPS spectra are unsuitable for
the measurement of discontinuity linestrengths such as D4000 (Bruzual
1993), since these are very sensitive to flux-calibration. Moreover,
our instrumental setup, especially at WIYN, has very poor response at
$\la$4000\AA.

To measure linestrengths, we chose to use the program {\sc INDEXF}
(Cardiel, Gorgas and Cenarro; see Cenarro et al. 2001) because of its
careful calculation of linestrength errors, following
\cite{Cardiel98}.  The estimated error spectrum (incorporating
Poisson-errors and a contribution from sky-subtraction noise) yields a
``photometric'' error, which is added in quadrature with the
(Monte-Carlo derived) errors due to the uncertainty in radial
velocity, yielding a total error due to photon noise, \ephot. 

The linestrengths are measured both at the $\sim$3 \AA\ native
resolution of the NFPS spectra and also at the $\sim$9 \AA\ resolution
of the Lick system.  The galaxy spectra were broadened to the Lick
resolution for each linestrength following the resolution curve from
\cite{Worthey97}.  For the NFPS H$\alpha$ linestrengths that lie
redward of the Lick spectral range, we smoothed by 10.5 \AA\ FWHM.
The Lick-resolution linestrengths are suitable for comparison to
synthesis models such as those of Worthey \citep{Worthey94}, based on
low-resolution stellar libraries. The full-resolution linestrengths in
principle retain more detailed information (at least for galaxies with
small velocity broadening), and can be analyzed in comparison to
higher-resolution synthesis models, such as those of
\cite{Vazdekis99}.

\subsection{Velocity broadening and aperture corrections}
\label{subsec:corrs}

Galaxy spectra are broadened by the line-of-sight velocities of their
stars, generally resulting in a dilution in measured linestrength with
increasing velocity dispersion, $\sigma$.  We followed the standard
procedure of determining correction curves from artificially broadened
K-giant stellar spectra.  The correction curve for the WIYN data is the average 
of the curves from nine stars while the CTIO correction curve is the average 
of the curves from twelve stars.  These corrections are very stable and
therefore we applied a common correction curve to data from all of the
observing runs.  For our Lick-resolution galaxy spectra, the stars
were first broadened to the Lick resolution before deriving the
correction curves.  Our corrections are very similar to those obtained
by other groups (e.g., Poggianti et al. 2001). 
The velocity broadening corrections derived from different template stars 
agree within $\lesssim$10\% for 
all indices except \hb.  However, in absolute 
terms the velocity broadening correction for \hb\ is quite small 
($\lesssim20\%$ of the index value).  

The NFPS spectra sample the galaxy light within a fixed 2\,arcsec
diameter.  This angular scale samples different physical scales for
galaxies at different angular diameter distances.  Because galaxies
have internal gradients, it is necessary to correct the raw data for
such aperture effects. In NFPS-I, we corrected the velocity
dispersions according to the prescription of \cite{Jorg95}.

Aperture corrections can also be applied to linestrengths to correct
for increasing apparent galaxy size at higher redshift.
We followed the formula
\begin{equation}
I_{\rm cor} = I_{\rm ap} + \Delta_{\rm ap}
\end{equation}
where $I_{\mathrm {ap}}$ is the uncorrected linestrength value
measured through a 2\,arcsec aperture, and $\Delta_{\rm ap}$ is given by 
\begin{equation}
\Delta_{\rm ap} = -\kappa\ \log\frac{d}{d_{0}}\,.
\end{equation}
The linestrength gradient
$\kappa = d\log({\mathrm{index}})/d\log(r_{\mathrm {ap}})$, 
and $\frac{d}{d_{0}}$ is a ratio of
angular diameter distances. Here $d_{0}$ is a normalization factor
defined as the angular diameter distance at $z = 0.05$.  For each
galaxy, $d$ is the angular diameter distance corresponding to the
cluster CMB-frame redshift; for non-cluster galaxies, ${d}$ is 
a function of 
the CMB-frame redshift of the galaxy itself.  In calculating ${d}$, we
adopted a cosmology with $\Omega_{m} = 0.3$ and $\Omega_{\Lambda} =
0.7$.  Thus the physical diameter of the corrected aperture diameter
is 1.37 \hkpc.  
Note that this correction is to a fixed metric diameter and not to a 
multiple of the effective radius, $R_e$. 
The gradients $\kappa$ were compiled from other groups who either
measured internal linestrength gradients at varying radii or extracted
multiple apertures for a galaxy, and are summarized in Table~\ref{tab:ap}.  
In particular, gradients either from \cite{Kunt02} or \cite{Proctor02}
were used for most of our linestrengths.  Table~\ref{tab:ap} also lists 
the mean difference between the corrected and uncorrected linestrengths for 
each line index, and the dispersion in this difference.  The difference 
between the corrected and uncorrected linestrengths varies by galaxy, but 
is generally at the level of a few percent.

\begin{deluxetable}{lcccc}
\tablecaption{Aperture Corrections}
\tablewidth{0pt}
\tablehead{
\colhead{Line Index}  &
\colhead{Units     }  &
\colhead{$\kappa$}  &
\colhead{$\langle\Delta_{\mathrm {ap}}\rangle$}  &
\colhead{$\langle\Delta_{\mathrm {ap}}^2\rangle^\half$}
} 
\startdata
H$\delta_{\rm A}$ & \AA\ & 1.283 & -0.071 & 0.172  \\
H$\delta_{\rm F}$ & \AA\ & 0.351 & -0.019 & 0.047  \\
CN$_{1}$ & mag & -0.058          & 0.003 & 0.008   \\
Ca4227 & \AA\ & -0.121           & 0.007 & 0.016   \\
H$\gamma_{\rm A}$ & \AA\ & 1.349 & -0.075 & 0.181  \\
H$\gamma_{\rm F}$ & \AA\ & 0.629 & -0.035 & 0.084  \\
Fe4383 & \AA\ & -1.484           & 0.082 & 0.199   \\
Ca4455 & \AA\ & -0.226           & 0.013 & 0.030   \\
Fe4531 & \AA\ & -0.544           & 0.030 & 0.073   \\
Fe4668 & \AA\ & -2.580           & 0.143 & 0.346   \\
\hb\ & \AA\ & 0.000              & 0.000 & 0.000   \\
Fe5015 & \AA\ & -1.346           & 0.075 & 0.180   \\
Mg$_{1}$ & mag & -0.040          & 0.003 & 0.007   \\
Mg$_{2}$ & mag & -0.066          & 0.005 & 0.011   \\
Mg{\it b} & \AA\ & -1.375        & 0.080 & 0.182   \\
Fe5270 & \AA\ & -0.706           & 0.091 & 0.095   \\
Fe5335 & \AA\ & -0.706           & 0.049 & 0.118   \\
Fe5406 & \AA\ & -0.460           & 0.014 & 0.062   \\
Fe5709 & \AA\ & -0.019           & 0.001 & 0.003   \\
Fe5782 & \AA\ & -0.019           & 0.001 & 0.002   \\
Na5895 & \AA\ & -0.045           & 0.006 & 0.006   \\
TiO$_{1}$ & mag & -0.016         & 0.002 & 0.003   \\
TiO$_{2}$ & mag & -0.011         & 0.001 & 0.001   \\
H$\alpha$A & \AA\ & 0.035        & -0.008 & 0.003  \\
H$\alpha$F & \AA\ & 0.035        & -0.008 & 0.004  \\
\enddata
\label{tab:ap}
\end{deluxetable}

The aperture corrections in velocity dispersion and linestrengths were
not applied for purposes of comparisons in
Sections~\ref{subsec:errors} and \ref{subsec:compare}.  However in
Section~\ref{sec:ixsig} and subsequent sections, the velocity
dispersions and linestrengths are corrected for aperture effects.

\subsection{Inter-run comparisons}
\label{subsec:errors}

As described above, we chose to measure our linestrengths with INDEXF
largely because of its method of calculating errors, which allows for
noise in the spectra as well as errors in the radial velocities.
However, given the possibility of systematic errors, it is useful to
check the consistency of the error estimates by comparing repeat
observations of the same galaxies.

Within the NFPS survey several ``standard'' equatorial clusters were
re-observed on several runs, and from each telescope.  The primary
goal of this was to ensure that velocity dispersions were on a
consistent base system and to correct discrepant runs if necessary. In
NFPS-I, we showed that velocity dispersions could be calibrated to a
systematic accuracy of $\sim$0.004 dex (thus limiting systematic
errors in FP-derived distances to $\le$1.5\%).

\begin{figure*}
\center{
\includegraphics[width=140mm,angle=270]{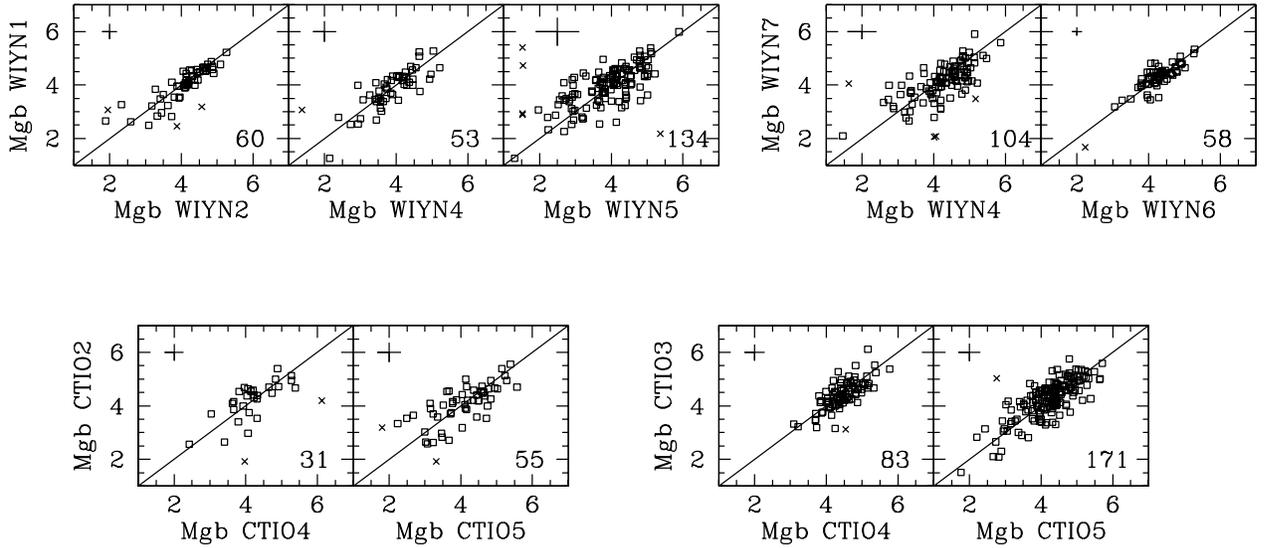}
}
\vskip -50mm
\caption{
Overlapping observations between CTIO runs and WIYN runs.  The line in
each plot has a slope of unity, representing where galaxies should
fall if there is no offset between multiple observations.  Crosses
indicate galaxies more than 4$\sigma$ away from the mean weighted
least squares fit between the two axes.  Average errors are shown in
the upper left of the plot and the number of galaxies is displayed in
the lower right.  Mg{\it b} is in units of angstroms and measured at
full (instrumental) resolution.
}
\label{bothcomp}
\end{figure*}

As an example of the consistency of our linestrength measurements, we
compare the Mg{\it b} instrumental-resolution absorption line
measurements in galaxies that were observed in multiple runs at the
same telescope and also at both sites.  In Figure~\ref{bothcomp} we
show comparisons of galaxies observed in more than one CTIO or WIYN
run.

While Poisson errors are the dominant source of error in the
linestrength index measurements, results for the same galaxy may also
differ due to other effects such as fiber position errors and
variations in fiber transmission.  Such errors will affect individual
measurements in a way that is likely to be independent of the
signal-to-noise ratio. To allow for these effects, we include an
``external error,'' \efib, assumed to be constant for all measurements
(but of course different from one linestrength to another), added in
quadrature with the estimated measurement errors, so that the total
random error $\eran = (\ephot^2+\efib^2)^{1/2}$. We determine the
value of this external error by requiring a good $\chi^2$ for repeated
measurements of the galaxies in common.  We find that for most
indices, the external errors are typically much smaller than the
typical random errors.  Nevertheless, it is necessary to account for
them since accurate characterization of measurement errors will be
important in Section~\ref{sec:trends} when we attempt to determine the
intrinsic population variation in linestrength.  For six indices
(Mg$_{1,2}$, TiO$_{1,2}$, H$\alpha_{\rm A}$, H$\alpha_{\rm F}$), the
external errors are slightly larger than the typical random error.  We
do not use any of these six indices to determine stellar population
parameters.

There is also the possibility of small systematic offsets in the
linestrength measurements between runs, due, for example, to
variations in seeing or to different instrumental configurations.  To
calculate additive corrections, $\Delta_{\mathrm {run}}$, needed to
bring the linestrength measurements from different runs onto a
common system, we use the repeat measurements and follow the same
procedure as was used for velocity dispersions in NFPS-I.  We find in
that in most cases (with the exception of the H$\alpha$ indices) the
run corrections are smaller than the typical random errors and are
very much smaller than the range spanned by the linestrength data.
Nevertheless, we prefer to include these small corrections so as not
to bias future bulk flow measurements derived from the FP relation
which also incorporate these linestrengths as additional parameters.
These corrections are applied to the raw linestrengths before the data
are merged to yield a final set of linestrength data for each galaxy. 
In Table~\ref{tab:systematics}, for each linestrength we list the total 
dispersion in the measured linestrengths for a given index, the typical 
photon error, \ephot, the systematic error, \efib, the total
random error, \eran, and the typical (root-mean-square) run correction
$\langle\Delta_{\rm run}^2\rangle^\half$.

\begin{deluxetable}{lccccc}
\tablecaption{Linestrength Index Errors}
\tablewidth{0pt}
\tablehead{
\colhead{Line Index} &
\colhead{ $\langle I^2\rangle^\half$\tablenotemark{a}} &
\colhead{\ephot\tablenotemark{b}} &
\colhead{\efib\tablenotemark{c}}  &
\colhead{\eran\tablenotemark{d}} &
\colhead{$\langle\Delta_{\rm run}^2\rangle^\half$\tablenotemark{e}}
}
\startdata
 H$\delta_{\rm A}$ &  2.14 & 0.91 & 0.58 & 0.99 & 0.30\\
 H$\delta_{\rm F}$ &  1.24 & 0.61 & 0.33 & 0.64 & 0.13\\
 CN$_{1}$          &  0.06 & 0.02 & 0.02 & 0.03 & 0.01\\
 CN$_{2}$          &  0.07 & 0.03 & 0.02 & 0.03 & 0.02\\
 Ca4227            &  0.60 & 0.40 & 0.20 & 0.42 & 0.06\\
 G4300             &  1.08 & 0.61 & 0.36 & 0.66 & 0.28\\
 H$\gamma_{\rm A}$ &  1.51 & 0.64 & 0.45 & 0.73 & 0.32\\
 H$\gamma_{\rm F}$ &  0.88 & 0.39 & 0.20 & 0.41 & 0.08\\
 Fe4383            &  1.15 & 0.75 & 0.37 & 0.78 & 0.26\\
 Ca4455            &  0.53 & 0.43 & 0.00 & 0.40 & 0.08\\
 Fe4531            &  0.77 & 0.53 & 0.25 & 0.55 & 0.09\\
 Fe4668            &  1.51 & 0.68 & 0.50 & 0.80 & 0.33\\
 \hb\              &  0.60 & 0.27 & 0.11 & 0.28 & 0.05\\
 \hb$_p$           &  0.52 & 0.18 & 0.09 & 0.20 & 0.02\\
 Fe5015            &  1.04 & 0.59 & 0.33 & 0.64 & 0.14\\
 Mg$_{1}$          &  0.03 & 0.01 & 0.01 & 0.01 & 0.01\\
 Mg$_{2}$          &  0.04 & 0.01 & 0.01 & 0.01 & 0.01\\
 Mg{\it b}         &  0.66 & 0.26 & 0.12 & 0.27 & 0.03\\
 Fe5270            &  0.44 & 0.28 & 0.11 & 0.28 & 0.04\\
 Fe5335            &  0.50 & 0.34 & 0.11 & 0.34 & 0.05\\
 Fe5406            &  0.35 & 0.25 & 0.11 & 0.26 & 0.04\\
 Fe5709            &  0.27 & 0.20 & 0.10 & 0.21 & 0.02\\
 Fe5782            &  0.27 & 0.17 & 0.10 & 0.19 & 0.05\\
 Na5895            &  0.95 & 0.18 & 0.16 & 0.23 & 0.08\\
 TiO$_{1}$         &  0.01 & 0.00 & 0.01 & 0.01 & 0.00\\
 TiO$_{2}$         &  0.01 & 0.00 & 0.01 & 0.01 & 0.00\\
 H$\alpha$A        &  1.23 & 0.15 & 0.15 & 0.21 & 0.31\\
 H$\alpha$F        &  2.67 & 0.14 & 0.19 & 0.24 & 0.09\\
\enddata
\label{tab:systematics}
\tablenotetext{a}{The total dispersion over the NFPS in the measured index, arising
  from intrinsic population differences and measurement error.}
\tablenotetext{b}{Median random error.}
\tablenotetext{c}{Mean systematic error.}
\tablenotetext{d}{Median total random error in merged sample.}
\tablenotetext{e}{Root-mean-square correction over all runs.}
\end{deluxetable}

\subsection{Comparisons with other surveys} 
\label{subsec:compare}
We compare our linestrengths with data from two other surveys:
$\sim$140 galaxies overlapping with the second data release from the
Sloan Digital Sky Survey \citep{SDSSdr2} and 33 Coma Cluster galaxies
from \cite{Moore02}.  The comparison with Moore et al. are at the Lick
resolution.  SDSS linestrengths available in their archive are at
their instrumental resolution (2.4 \AA) while our instrumental
resolution linestrengths are at 3 \AA.  Using the ratio of our Lick
resolution linestrengths to our full resolution linestrengths, we
``scaled'' the SDSS linestrengths to 3 \AA\ resolution.  The
correction to each linestrength is typically very small ($\sim$2\%).
None of the data sets have aperture corrections applied for these
comparisons.

In Table~\ref{tab:compare} we compare our measurements with those from
other surveys, for several linestrengths in common.  The mean offsets
are defined as $\Delta = <I{\rm _{NFPS}}$ - $I{\rm_{other}}>$ where
$I{\rm _{NFPS}}$ is the NFPS linestrength and $I{\rm_{other}}$ is the
SDSS or Moore et al. linestrength, and the rms is the standard
deviation of the differences.

\begin{deluxetable}{lccccc}
\tablecaption{Line Index Offsets}
\tablewidth{0pt}
\tablehead{
\colhead{Survey}  & 
\colhead{Line Index}   &
\colhead{Units}   &
\colhead{$N_{\rm gals}$}    &
\colhead{  Mean offset}   &
\colhead{rms}}
\startdata
Moore et al & Fe4668   & \AA\ & 34 &  0.236$\pm$0.174 & 0.694 \\
	    & \hb\     & \AA\ & 34 & -0.035$\pm$0.034 & 0.625 \\
	    & Fe5015   & \AA\ & 34 &  0.186$\pm$0.119 & 0.769 \\
	    & Mg$_{1}$ & mag  & 34 &  0.017$\pm$0.002 & 0.019 \\
	    & Mg$_{2}$ & mag  & 34 & -0.005$\pm$0.003 & 0.030 \\
	    & Mg{\it b}& \AA\ & 34 & -0.079$\pm$0.036 & 0.473 \\
	    & Fe5270   & \AA\ & 34 & -0.039$\pm$0.058 & 0.420 \\
	    & Fe5335   & \AA\ & 34 &  0.063$\pm$0.049 & 0.340 \\ 
\\
SDSS        & Ca4227   & \AA\ & 141 & 0.102$\pm$0.039 & 0.468 \\ 
	    & Fe4383   & \AA\ & 141 & 0.261$\pm$0.096 & 1.139 \\
	    & Ca4455   & \AA\ & 141 & 0.271$\pm$0.046 & 0.543 \\ 
	    & Fe4531   & \AA\ & 141 & 0.231$\pm$0.092 & 1.090 \\
	    & Fe4668   & \AA\ & 141 & -0.046$\pm$0.115 & 1.362 \\
	    & \hb\     & \AA\ & 141 & 0.021$\pm$0.045 & 0.540 \\
	    & Fe5015   & \AA\ & 140 & 0.601$\pm$0.078 & 0.920 \\
	    & Mg{\it b}& \AA\ & 136 & 0.126$\pm$0.039 & 0.455 \\
	    & Fe5270   & \AA\ &  64 & 0.220$\pm$0.066 & 0.529 \\
	    & Fe5335   & \AA\ & 102 & 0.406$\pm$0.054 & 0.547 \\
	    & Fe5406   & \AA\ & 135 & 0.208$\pm$0.037 & 0.431 \\
	    & Fe5709   & \AA\ & 129 & 0.095$\pm$0.032 & 0.366 \\
	    & Fe5782   & \AA\ &  84 & 0.021$\pm$0.031 & 0.287 \\
	    & Na5895   & \AA\ &  71 & 0.083$\pm$0.052 & 0.437 \\
\enddata
\label{tab:compare}
\end{deluxetable}

\begin{figure}
\center{
\includegraphics[height=70mm,width=90mm,angle=0]{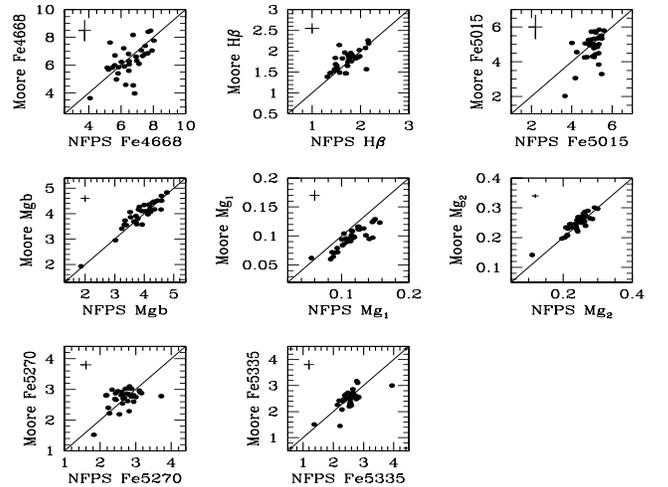}
}
\vskip -1mm
\caption{NFPS linestrengths for several line indices
plotted against those from \cite{Moore02}.  Solid lines with a slope
of unity indicate exact agreement between the two sets of
measurements.  All measurements are at Lick resolution.  The average
NFPS and Moore et al. linestrength error is shown in the upper left
corner of each plot.}
\label{fig:moore}
\end{figure}

We plot the NFPS linestrengths against those of Moore et al. in
Figure~\ref{fig:moore}.  Our data are in very good agreement with
those of Moore et al.; with the exception of Mg$_1$, which, as noted
above, is susceptible to flux calibration error, there is no evidence
for systematic offsets. The $\chi^2$ values of the comparisons are
acceptable for all linestrengths except for Fe4668, Fe5270 and Mg$_2$.
This suggests that, in most cases, both NFPS and Moore et al.'s errors
are reasonable.

\begin{figure*}
\center{
\includegraphics[height=160mm,width=180mm,angle=0]{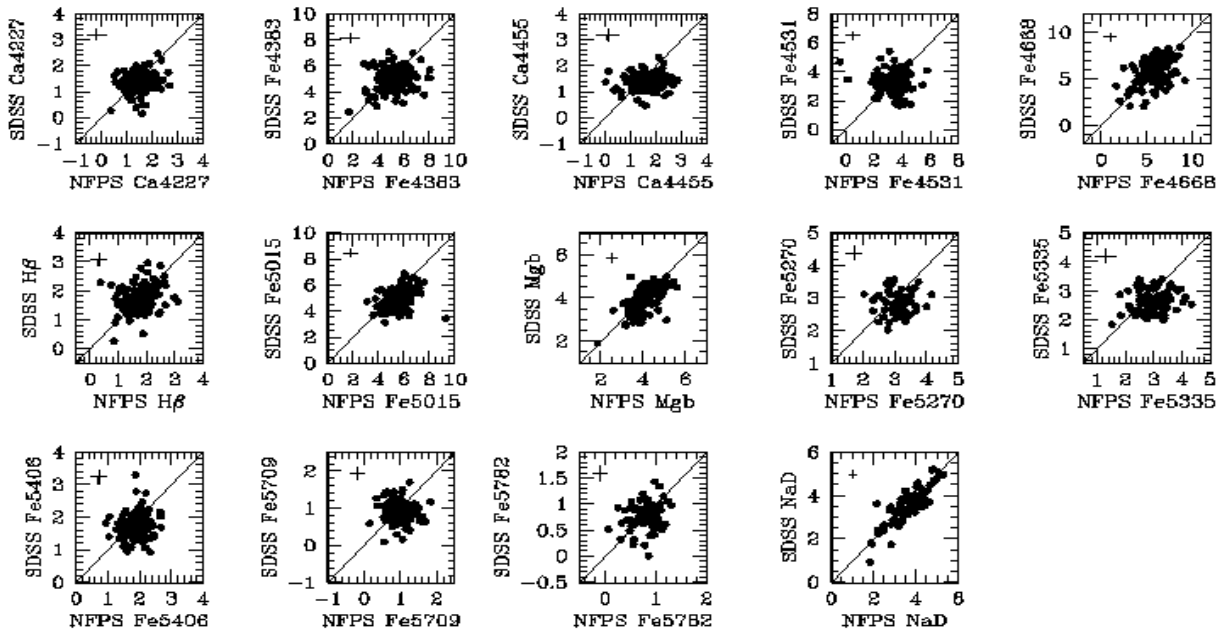}
}
\vskip -65mm
\caption{Plots of NFPS linestrengths at instrumental
resolution against SDSS linestrengths ``scaled'' to the same NFPS
resolution.  Solid lines with a slope of unity indicate exact
agreement between the two sets of measurements.  The average NFPS and
SDSS linestrength error is shown in the upper left corner of each
plot.}
\label{fig:sdss}
\end{figure*}

The comparison with SDSS is shown in Figure~\ref{fig:sdss}.  Note that
for linestrengths in the blue, the quoted SDSS errors are smaller,
whereas for linestrengths in the red, the NFPS errors are typically
smaller.  There is evidence for small but significant offsets for most
of the linestrengths, as indicated in
Table~\ref{tab:compare}. Furthermore, for most linestrengths, the
reduced $\chi^2$ values are larger than unity, suggesting that the
errors in either SDSS or NFPS (or both) are underestimated. 
Overall, however, based on the consistency between our measurements
and those of Moore et al. we conclude that our error estimates, which
include systematic effects estimated from repeat observations, are
realistic.

\section{Nebular Emission Measurements}
\label{sec:emiss}

While elliptical galaxy spectra are broadly characterized as being
dominated by absorption lines, it has long been realized that many
systems also show nebular emission lines
\citep{Mayall57,Phillips86,Gonzalez93,Goudfrooij94}.

A particular difficulty in studying integrated stellar populations is
that nebular \hb\ emission acts to `fill in' the stellar absorption
line, driving derived age measurements toward older
values. Disentangling the absorption and emission components of \hb\
is therefore vital if this linestrength is to be used to constrain
star-formation histories.  Since emission at \hb\ is often co-incident
with emission in \oiiiab, a standard approach is to establish a
correction based on the more easily-measured \oiii\ lines, usually
assuming a constant correction factor of \hb$=0.6\times$\oiiib
\citep{Trager00a}.  In this section, we discuss an alternative method
which aims to distinguish whether galaxies are likely affected by
emission by fitting the spectra with absorption template models.

\subsection{Method}

Since galaxy spectra contain substantial structure from unresolved
stellar absorption lines, weak \oiii\ lines can be very difficult to
distinguish without first removing the stellar component. This can be
done by subtracting/dividing a `matched' spectrum for similar galaxies
of similar stellar population properties but free from emission
\citep{Goudfrooij94} or by subtracting/dividing a model stellar
continuum \citep{Kunt02}. In the latter case it is necessary to
exclude the region around Mg$b$ where current synthetic spectra (such
as those of Vazdekis 1999) provide a poor match to observed elliptical
galaxies due to non-solar abundance ratios.  This limits the nebular
lines which can be measured, excluding such potentially-interesting
features as N\,I$\lambda5199$, but the method has the advantage of
being easily automated when the redshift and velocity dispersion are
known.

\begin{figure*}
\includegraphics[width=130mm,angle=270]{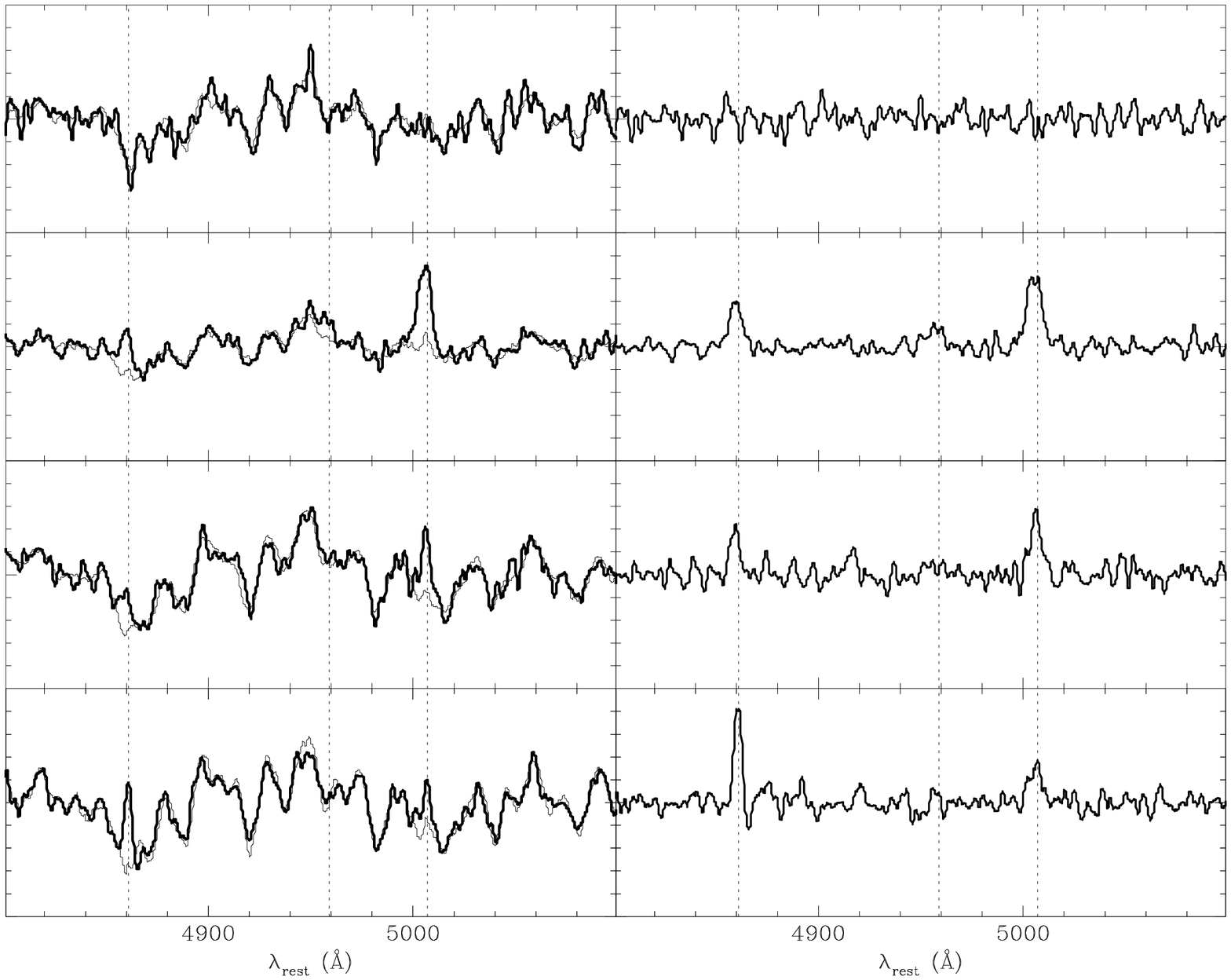}
\vskip -7mm
\caption{Some illustrative examples of the emission-line
measurements.  In the left-hand panels are shown the observed spectrum
(heavy line) and the best-fit Vazdekis (1999) model (thin line). The
ratio spectrum, from which the line measurements are derived, is shown
in the corresponding right-hand panel. Vertical dotted-lines indicate
the expected location of the \hb\ and \oiii\ lines. At the top, a
spectrum with no measurable emission; in the lower three panels,
spectra with weak-to-moderate emission.  Note the substantial range of
\hb\ to \oiii\ ratios among these examples.}
\label{fig:emissexams}
\end{figure*}

Our measurements of \hb\ and \oiii\ emission are made on the spectra
after dividing by the best-fitting \cite{Vazdekis99} model, computed
over the $4800-5100$\,\AA\ spectral range. The models span a range of
metallicity $-0.7<$[Fe/H]$<+0.2$ and age $1.0\,$Gyr$ < t < 17.4\,$Gyr,
and assume solar abundance ratios and power-law IMF with $x=1.3$.  The
best-fit model is computed by comparing the model and observed spectra
after first shifting and broadening the model to match the redshift
and velocity dispersion of the galaxy, as determined in Paper I.
After division by the best-fitting model, a low-order continuum is
divided out to remove long-wavelength baseline variations. Equivalent
widths are measured directly on the divided spectra (i.e., without
assuming an emission-line profile), and errors estimated based on the
noise in the `line-free' regions. Figure~\ref{fig:emissexams} shows
some illustrative examples of this continuum-removal method.

It is clear that the model-fitting procedure itself can be biased by
the presence of \hb\ emission. Specifically, for galaxies with
emission, a model with weaker stellar \hb\ can be fitted, yielding an
underestimate of the nebular \hb. We have investigated this effect
through simulations, adding emission lines to model population spectra
of varying ages and velocity broadening. The recovered \hb\ are biased
low by 10--25\%, depending on the underlying spectrum.  In spectra of
high velocity dispersion, the broad wings of the absorption spectra
help to distinguish the narrow nebular emission lines.  In older
populations, the bias is somewhat reduced because there is little
flexibility to push the fit to even older models. Thus the worst cases
(where emission \hb\ is underestimated by 25\%) are for low-$\sigma$
young stellar population spectra.  Despite this bias, the wings
of the \hb\ feature are essential to separate the broad stellar and
narrow nebular contributions.

\subsection{Results}
\label{sec:emiss_results}

The above method was applied to 4964 galaxies with redshift and
velocity dispersion data.  (For this purpose we allowed the use of
velocity dispersions from lower signal-to-noise spectra not reported
in NFPS-I.)  Of these, significant \oiiib\ emission is detected in 589
galaxies, EW(\oiiia) in 154 galaxies, and EW(\hb) in 633 galaxies (all
$3\sigma$ detection limits).

\begin{figure*}
\includegraphics[width=140mm,angle=270]{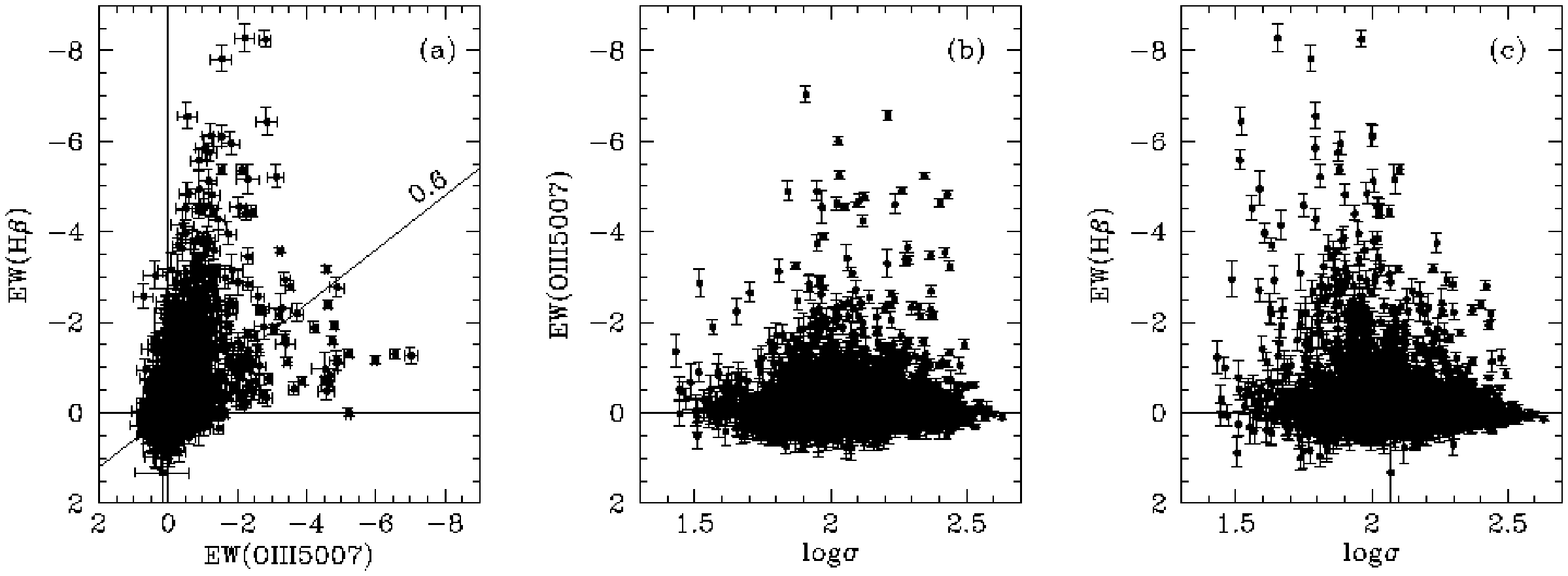}
\vskip -65mm 
\caption{Emission measurements at \oiiib\ and \hb\ as a
function of velocity dispersion, $\sigma$.}
\label{emiss_sigma}
\end{figure*}

The correlations between EW(\oiiib), EW(\hb), and $\sigma$ are
presented in Figure~\ref{emiss_sigma}.  From Figure 5a it is clear 
that while the presence of OIII is indeed an indicator for \hb\
emission, no single ratio between the lines is appropriate for all
galaxies. In particular, there is a substantial population of objects
with moderate to strong \hb\ emission but with little or no
\oiii. Plotting the emission lines versus velocity dispersion (Figures 5b,c), 
shows that these objects are overwhelmingly of low mass
($\log\sigma\la2.0$).  Thus for low-$\sigma$ objects, the classical
factor 0.6 under-corrects for emission contamination, while for
high-$\sigma$ objects, the factor over-corrects.  Note that the
objects with large \hb\ emission are those for which we expect in fact
to {\it underestimate} the measurement, due to the bias described
above.

\begin{figure}
\includegraphics[width=70mm,angle=0]{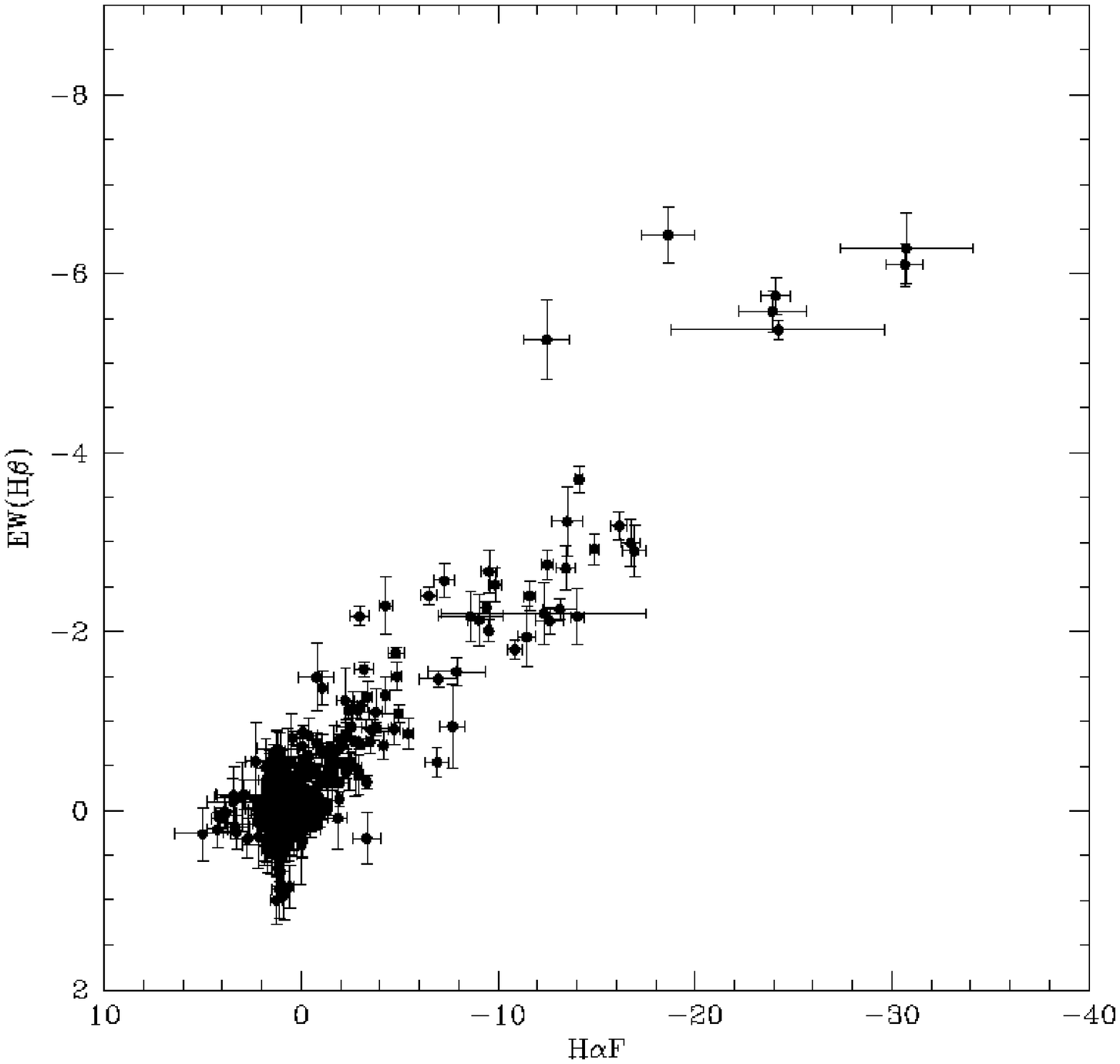}
\vskip -1mm
\caption{Emission-line measurements, versus the \haf\ linestrength 
(see Section~\ref{sec:emiss_results}).  Units are in \AA.}
\label{emiss_halpha}
\end{figure}

A better predictor of emission at \hb\ is the emission at
H$\alpha$ \citep{Caldwell03}.  As described in
Section~\ref{subsec:rawix}, \haf\ linestrength measurements are only
available for a subset of $\sim$700 galaxies, but these can provide a
consistency check on \hb\ emission, as shown in
Figure~\ref{emiss_halpha}.  The \haf\ linestrength is constructed so
as to be insensitive to the neighboring [NII] lines.  While \haf\ does
of course have some contribution from stellar absorption, the ratio
between the nebular emission lines is such that emission, when
present, usually dominates at \ha, even if it is weak at \hb.  The
excellent correlation in the upper panel suggests that the emission
\hb\ measurements are indeed at least a reasonable proxy for \ha.  The
\ha/\hb\ ratio ranges from 2.9 to 5.9 (68\% range), with a median of
4.5, similar to the typical range of 2--6 for early-type spirals
\citep{Stasi04}.

\subsection{Emission-line rejection criteria}

\begin{figure}
\includegraphics[height=70mm,width=70mm,angle=0]{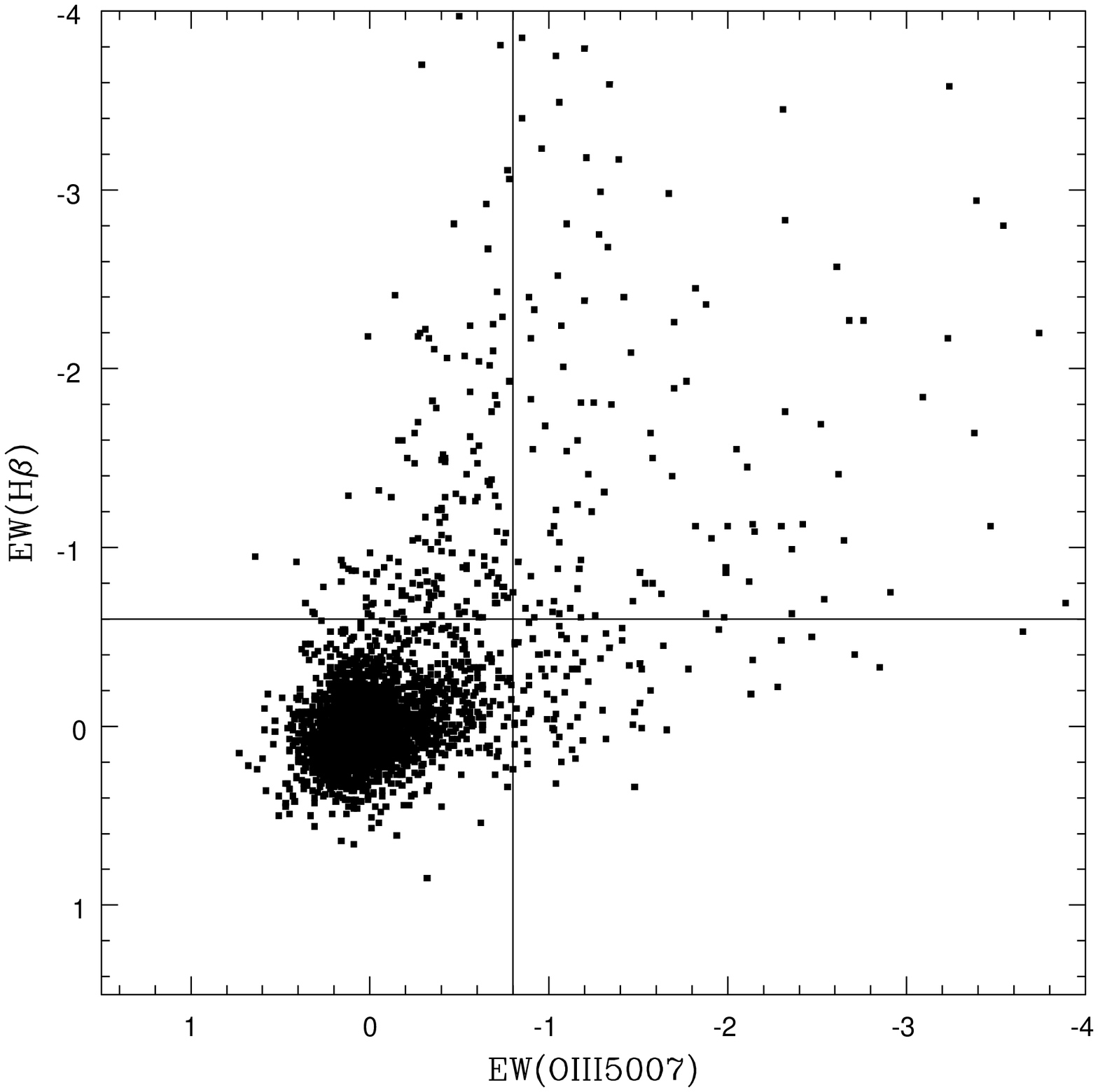}
\vskip -1mm
\caption{Equivalent widths of OIII$\lambda$5007 and \hb\ emission lines.
Solid lines are lines of constant emission.  Points above the
horizontal line and right of the vertical line are galaxies excluded
from our {\sc CULL} dataset.  Units are in \AA.}
\label{emiss}
\end{figure}

For galaxies with excess emission in the \hb\ and
OIII$\lambda$4959 and 5007\AA\ lines, the \hb\ absorption line index
is contaminated and hence the derived stellar properties are biased.
As described above, a constant correction factor is incorrect, even in
a statistical sense, since the true ratios vary widely and {\it
systematically} with velocity dispersion.  Therefore, instead of using
the constant correction factor based only on [OIII], we prefer simply
to exclude galaxies with emission at either \hb\ or \oiiib.  In
Figure~\ref{emiss} we plot the emission measurements of \hb\
against [OIII$\lambda$5007\AA] and indicate the cuts made to exclude
high-emission galaxies.  Only galaxies with EW(\hb) $> -0.6$ and
EW(OIII) $> -0.8$ are included in our sample of all linestrengths used
for analysis in the next two sections.

\section{Data Presentation and Summary}
\label{sec:sample}

Tables~\ref{tab:lickindexdata} and \ref{tab:fullindexdata} present
absorption and emission linestrengths for all galaxies observed,
including those with emission.  Hereafter, we refer to this as the
{\sc ALL} dataset.  Table~\ref{tab:lickindexdata} gives linestrengths
and errors at Lick resolution for galaxies with $S/N > $15 per
Angstrom, at 5000--5500 \AA.  The linestrengths in the table are
corrected for velocity broadening, but have not been corrected for
aperture effects.  Additionally, the OIII and \hb\ emission equivalent
widths with their common error are listed (indicated by $W$), along
with the heliocentric redshift (cz$_{\odot}$), velocity dispersion
($\sigma$), cluster identification and ratio of the angular diameter
distances based on CMB-frame redshifts. Table~\ref{tab:fullindexdata}
presents the full-resolution data in analogous form.  The auxiliary
parameters (cz$_{\odot}$, etc) are duplicated here for convenience.

\begin{table*}
\caption{Lick resolution linestrength indices and associated data}
\label{tab:lickindexdata}
\scriptsize
\begin{tabular}{cccccccccc}
\tableline
\tableline
\multicolumn{3}{c}{{\bf Galaxy ID}} & \ \ \ H$\delta_{\rm F}$ & \ \ \ H$\delta_{\rm A}$ & \ \ \ CN$_1$ & \ \ \ CN$_2$ & \ \ \ Ca4227 & \ \ \ G4300 & \ \ \ H$\gamma_{\rm F}$ \\
W(OIII) & W(H$\beta)$ & $\varepsilon$(W)& \ \ \ H$\gamma_{\rm A}$ & \ \ \ Fe4383 & \ \ \ Ca4455 & \ \ \ Fe4531 & \ \ \ Fe4668 & \ \ \ H$\beta$ & \ \ \ H$\beta^+$ \\
$cz_\odot$ & \multicolumn{2}{c}{$\log\sigma$} & \ \ \ Fe5015 & \ \ \ Mg$b$ & \ \ \ Mg$_1$ & \ \ \ Mg$_2$ & \ \ \ Fe5270 & \ \ \ Fe5335 & \ \ \ Fe5406 \\
\multicolumn{2}{l}{Cluster} & $\log(d/d_0)$ & \ \ \ Fe5709 & \ \ \ Fe5782 & \ \ \ Na5895 & \ \ \ TiO$_1$ & \ \ \ TiO$_2$ & \ \ \ H$\alpha_{\rm A}$ & \ \ \ H$\alpha_{\rm F}$ \\
\tableline
\tableline
\multicolumn{3}{l}{{\bf NFPJ001006.3-284623}} & $ \phantom{+}1.687\pm 0.748$& $ -1.355\pm 1.495$& $ \phantom{+}0.0266\pm 0.0416$& $ \phantom{+}0.0574\pm 0.0382$& $ \phantom{+}0.016\pm 0.698$& $ \phantom{+}6.174\pm 0.881$& $ -0.648\pm 0.673$\\
$-0.06$ & $-0.35$ & $\phantom{+}0.22$ & $ -4.244\pm 1.202$& $ \phantom{+}4.579\pm 1.062$& $ \phantom{+}0.674\pm 0.582$& $ \phantom{+}3.540\pm 0.852$& $ \phantom{+}4.268\pm 1.415$& $ \phantom{+}2.068\pm 0.459$& $ \phantom{+}2.342\pm 0.348$\\
18989 & \multicolumn{2}{c}{$1.877\pm0.050$}& $ \phantom{+}4.472\pm 0.921$& $ \phantom{+}3.051\pm 0.442$& $ \phantom{+}0.0435\pm 0.0133$& $ \phantom{+}0.1643\pm 0.0170$& \nodata& $ \phantom{+}2.884\pm 0.504$& $ \phantom{+}1.004\pm 0.411$\\
\multicolumn{2}{l}{A2734} & 0.080 & $ \phantom{+}1.499\pm 0.365$& \nodata& \nodata& \nodata& \nodata& \nodata& \nodata\\
\tableline
\multicolumn{3}{l}{{\bf NFPJ001010.7-285020}} & $ \phantom{+}0.580\pm 0.553$& $ -1.020\pm 1.001$& $ \phantom{+}0.0192\pm 0.0306$& $ \phantom{+}0.0446\pm 0.0345$& $ \phantom{+}1.149\pm 0.421$& $ \phantom{+}4.911\pm 0.675$& $ -0.361\pm 0.454$\\
$\phantom{+}0.03$ & $-0.13$ & $\phantom{+}0.14$ & $ -2.944\pm 0.836$& $ \phantom{+}3.607\pm 0.760$& $ \phantom{+}1.065\pm 0.336$& $ \phantom{+}1.936\pm 0.593$& $ \phantom{+}4.855\pm 0.966$& $ \phantom{+}2.126\pm 0.334$& $ \phantom{+}2.341\pm 0.239$\\
18626 & \multicolumn{2}{c}{$2.041\pm0.039$}& $ \phantom{+}4.730\pm 0.713$& $ \phantom{+}3.115\pm 0.351$& $ \phantom{+}0.0600\pm 0.0117$& $ \phantom{+}0.1609\pm 0.0151$& \nodata& $ \phantom{+}2.317\pm 0.380$& $ \phantom{+}1.687\pm 0.264$\\
\multicolumn{2}{l}{A2734} & 0.080 & $ \phantom{+}0.505\pm 0.239$& \nodata& \nodata& \nodata& \nodata& \nodata& \nodata\\
\tableline
\multicolumn{3}{l}{{\bf NFPJ001024.3-284935}} & $ \phantom{+}0.272\pm 0.645$& $ -1.753\pm 0.988$& $ \phantom{+}0.0553\pm 0.0267$& $ \phantom{+}0.0812\pm 0.0330$& $ \phantom{+}1.736\pm 0.448$& $ \phantom{+}4.635\pm 0.766$& $ -1.225\pm 0.479$\\
$\phantom{+}0.10$ & $\phantom{+}0.41$ & $\phantom{+}0.17$ & $ -5.047\pm 0.870$& $ \phantom{+}4.263\pm 0.895$& $ \phantom{+}1.193\pm 0.443$& $ \phantom{+}3.046\pm 0.622$& $ \phantom{+}4.245\pm 1.072$& $ \phantom{+}1.909\pm 0.363$& $ \phantom{+}2.030\pm 0.245$\\
17923 & \multicolumn{2}{c}{$2.234\pm0.041$}& $ \phantom{+}5.946\pm 0.764$& $ \phantom{+}3.599\pm 0.390$& $ \phantom{+}0.0563\pm 0.0119$& $ \phantom{+}0.2086\pm 0.0156$& \nodata& $ \phantom{+}3.133\pm 0.471$& $ \phantom{+}1.768\pm 0.346$\\
\multicolumn{2}{l}{A2734} & 0.080 & $ \phantom{+}0.998\pm 0.279$& \nodata& \nodata& \nodata& \nodata& \nodata& \nodata\\
\tableline
\multicolumn{3}{l}{{\bf NFPJ001032.5-285154}} & $ \phantom{+}0.309\pm 0.781$& $ -0.877\pm 1.168$& $ \phantom{+}0.0538\pm 0.0326$& $ \phantom{+}0.0830\pm 0.0421$& $ \phantom{+}0.912\pm 0.502$& $ \phantom{+}5.587\pm 0.799$& $ -1.787\pm 0.581$\\
$-0.17$ & $-0.04$ & $\phantom{+}0.19$ & $ -6.479\pm 1.041$& $ \phantom{+}5.265\pm 1.049$& $ \phantom{+}1.230\pm 0.439$& $ \phantom{+}4.227\pm 0.723$& $ \phantom{+}6.701\pm 1.138$& $ \phantom{+}1.282\pm 0.383$& $ \phantom{+}1.531\pm 0.316$\\
17441 & \multicolumn{2}{c}{$2.156\pm0.047$}& $ \phantom{+}5.538\pm 0.898$& $ \phantom{+}4.494\pm 0.422$& $ \phantom{+}0.1003\pm 0.0129$& $ \phantom{+}0.2501\pm 0.0168$& \nodata& $ \phantom{+}2.553\pm 0.476$& $ \phantom{+}1.312\pm 0.316$\\
\multicolumn{2}{l}{A2734} & 0.080 & $ \phantom{+}1.051\pm 0.284$& \nodata& \nodata& \nodata& \nodata& \nodata& \nodata\\
\tableline
\multicolumn{3}{l}{{\bf NFPJ001041.8-283444}} & $ \phantom{+}0.032\pm 1.019$& $ -2.041\pm 1.502$& $ \phantom{+}0.0132\pm 0.0413$& $ \phantom{+}0.0593\pm 0.0358$& $ \phantom{+}1.862\pm 0.589$& $ \phantom{+}4.676\pm 0.892$& $ -2.127\pm 0.696$\\
$-0.08$ & $\phantom{+}0.01$ & $\phantom{+}0.21$ & $ -5.915\pm 1.202$& $ \phantom{+}2.918\pm 1.555$& $ \phantom{+}0.692\pm 0.527$& $ \phantom{+}3.183\pm 0.767$& $ \phantom{+}3.673\pm 1.397$& $ \phantom{+}2.344\pm 0.511$& $ \phantom{+}2.435\pm 0.349$\\
18190 & \multicolumn{2}{c}{$1.988\pm0.053$}& $ \phantom{+}4.396\pm 0.987$& $ \phantom{+}2.885\pm 0.527$& $ \phantom{+}0.0140\pm 0.0135$& $ \phantom{+}0.1603\pm 0.0179$& \nodata& $ \phantom{+}2.059\pm 0.583$& $ \phantom{+}1.595\pm 0.376$\\
\multicolumn{2}{l}{A2734} & 0.080 & $ \phantom{+}0.436\pm 0.300$& \nodata& \nodata& \nodata& \nodata& \nodata& \nodata\\
\tableline
\multicolumn{3}{l}{{\bf NFPJ001042.6-284916}} & $ -0.028\pm 0.723$& $ -1.420\pm 1.070$& $ \phantom{+}0.0459\pm 0.0282$& $ \phantom{+}0.0712\pm 0.0332$& $ \phantom{+}1.368\pm 0.443$& $ \phantom{+}4.639\pm 0.756$& $ -0.680\pm 0.515$\\
$-0.60$ & $-0.80$ & $\phantom{+}0.20$ & $ -3.697\pm 0.933$& $ \phantom{+}4.181\pm 0.960$& $ \phantom{+}0.949\pm 0.415$& $ \phantom{+}2.020\pm 0.796$& $ \phantom{+}6.930\pm 1.133$& $ \phantom{+}1.633\pm 0.415$& $ \phantom{+}1.843\pm 0.309$\\
18682 & \multicolumn{2}{c}{$1.942\pm0.047$}& $ \phantom{+}4.511\pm 0.903$& $ \phantom{+}3.647\pm 0.454$& $ \phantom{+}0.0510\pm 0.0129$& $ \phantom{+}0.2026\pm 0.0168$& \nodata& $ \phantom{+}1.500\pm 0.472$& $ \phantom{+}1.817\pm 0.370$\\
\multicolumn{2}{l}{A2734} & 0.080 & $ \phantom{+}0.760\pm 0.287$& \nodata& \nodata& \nodata& \nodata& \nodata& \nodata\\
\tableline
\tableline
\end{tabular}
Table~\ref{tab:lickindexdata} is presented in its entirety in the electronic 
edition of the Astrophysical Journal. A portion is shown here for guidance 
regarding its form and content.
\end{table*}

\begin{table*}
\caption{Full resolution linestrength indices and associated data}
\label{tab:fullindexdata}
\scriptsize
\begin{tabular}{cccccccccc}
\tableline
\tableline
\multicolumn{3}{c}{{\bf Galaxy ID}} & \ \ \ H$\delta_{\rm F}$ & \ \ \ H$\delta_{\rm A}$ & \ \ \ CN$_1$ & \ \ \ CN$_2$ & \ \ \ Ca4227 & \ \ \ G4300 & \ \ \ H$\gamma_{\rm F}$ \\
W(OIII) & W(H$\beta)$ & $\varepsilon$(W)& \ \ \ H$\gamma_{\rm A}$ & \ \ \ Fe4383 & \ \ \ Ca4455 & \ \ \ Fe4531 & \ \ \ Fe4668 & \ \ \ H$\beta$ & \ \ \ H$\beta^+$ \\
$cz_\odot$ & \multicolumn{2}{c}{$\log\sigma$} & \ \ \ Fe5015 & \ \ \ Mg$b$ & \ \ \ Mg$_1$ & \ \ \ Mg$_2$ & \ \ \ Fe5270 & \ \ \ Fe5335 & \ \ \ Fe5406 \\
\multicolumn{2}{l}{Cluster} & $\log(d/d_0)$ & \ \ \ Fe5709 & \ \ \ Fe5782 & \ \ \ Na5895 & \ \ \ TiO$_1$ & \ \ \ TiO$_2$ & \ \ \ H$\alpha_{\rm A}$ & \ \ \ H$\alpha_{\rm F}$ \\
\tableline
\tableline
\multicolumn{3}{l}{{\bf NFPJ001006.3-284623}} & $ \phantom{+}2.186\pm 0.977$& $ -1.756\pm 1.561$& $ \phantom{+}0.0355\pm 0.0418$& $ \phantom{+}0.0824\pm 0.0493$& $ \phantom{+}0.141\pm 0.692$& $ \phantom{+}7.007\pm 1.039$& $ -0.874\pm 0.703$\\
$-0.06$ & $-0.35$ & $\phantom{+}0.22$ & $ -4.250\pm 1.221$& $ \phantom{+}5.697\pm 1.330$& $ \phantom{+}0.885\pm 0.755$& $ \phantom{+}3.795\pm 0.909$& $ \phantom{+}4.022\pm 1.343$& $ \phantom{+}2.189\pm 0.499$& $ \phantom{+}2.211\pm 0.338$\\
18989 & \multicolumn{2}{c}{$1.877\pm0.050$}& $ \phantom{+}4.972\pm 1.033$& $ \phantom{+}3.213\pm 0.475$& $ \phantom{+}0.0448\pm 0.0129$& $ \phantom{+}0.1671\pm 0.0164$& \nodata& $ \phantom{+}3.212\pm 0.559$& $ \phantom{+}1.016\pm 0.434$\\
\multicolumn{2}{l}{A2734} & 0.080 & $ \phantom{+}1.333\pm 0.338$& \nodata& \nodata& \nodata& \nodata& \nodata& \nodata\\
\tableline
\multicolumn{3}{l}{{\bf NFPJ001010.7-285020}} & $ \phantom{+}0.703\pm 0.689$& $ -1.443\pm 1.079$& $ \phantom{+}0.0275\pm 0.0306$& $ \phantom{+}0.0515\pm 0.0366$& $ \phantom{+}1.345\pm 0.498$& $ \phantom{+}5.296\pm 0.783$& $ -0.446\pm 0.491$\\
$\phantom{+}0.03$ & $-0.13$ & $\phantom{+}0.14$ & $ -3.011\pm 0.850$& $ \phantom{+}4.431\pm 0.941$& $ \phantom{+}1.551\pm 0.479$& $ \phantom{+}2.328\pm 0.701$& $ \phantom{+}5.154\pm 1.025$& $ \phantom{+}2.185\pm 0.361$& $ \phantom{+}2.322\pm 0.246$\\
18626 & \multicolumn{2}{c}{$2.041\pm0.039$}& $ \phantom{+}5.426\pm 0.822$& $ \phantom{+}3.179\pm 0.371$& $ \phantom{+}0.0609\pm 0.0110$& $ \phantom{+}0.1617\pm 0.0143$& \nodata& $ \phantom{+}2.829\pm 0.456$& $ \phantom{+}2.094\pm 0.343$\\
\multicolumn{2}{l}{A2734} & 0.080 & $ \phantom{+}0.546\pm 0.274$& \nodata& \nodata& \nodata& \nodata& \nodata& \nodata\\
\tableline
\multicolumn{3}{l}{{\bf NFPJ001024.3-284935}} & $ \phantom{+}0.243\pm 0.677$& $ -2.122\pm 1.066$& $ \phantom{+}0.0744\pm 0.0312$& $ \phantom{+}0.1053\pm 0.0367$& $ \phantom{+}1.929\pm 0.507$& $ \phantom{+}4.841\pm 0.850$& $ -1.280\pm 0.514$\\
$\phantom{+}0.10$ & $\phantom{+}0.41$ & $\phantom{+}0.17$ & $ -4.579\pm 0.880$& $ \phantom{+}4.710\pm 1.004$& $ \phantom{+}1.509\pm 0.555$& $ \phantom{+}3.436\pm 0.706$& $ \phantom{+}4.078\pm 1.048$& $ \phantom{+}1.934\pm 0.385$& $ \phantom{+}2.055\pm 0.258$\\
17923 & \multicolumn{2}{c}{$2.234\pm0.041$}& $ \phantom{+}7.011\pm 0.902$& $ \phantom{+}3.674\pm 0.409$& $ \phantom{+}0.0589\pm 0.0114$& $ \phantom{+}0.2105\pm 0.0149$& \nodata& $ \phantom{+}3.641\pm 0.543$& $ \phantom{+}1.911\pm 0.390$\\
\multicolumn{2}{l}{A2734} & 0.080 & $ \phantom{+}1.067\pm 0.310$& \nodata& \nodata& \nodata& \nodata& \nodata& \nodata\\
\tableline
\multicolumn{3}{l}{{\bf NFPJ001032.5-285154}} & $ \phantom{+}0.490\pm 0.809$& $ -1.176\pm 1.242$& $ \phantom{+}0.0668\pm 0.0372$& $ \phantom{+}0.0923\pm 0.0440$& $ \phantom{+}1.185\pm 0.646$& $ \phantom{+}6.150\pm 0.924$& $ -1.862\pm 0.612$\\
$-0.17$ & $-0.04$ & $\phantom{+}0.19$ & $ -6.273\pm 1.036$& $ \phantom{+}5.713\pm 1.152$& $ \phantom{+}1.811\pm 0.637$& $ \phantom{+}4.644\pm 0.800$& $ \phantom{+}6.825\pm 1.166$& $ \phantom{+}1.477\pm 0.457$& $ \phantom{+}1.438\pm 0.308$\\
17441 & \multicolumn{2}{c}{$2.156\pm0.047$}& $ \phantom{+}6.469\pm 1.051$& $ \phantom{+}4.668\pm 0.449$& $ \phantom{+}0.1047\pm 0.0125$& $ \phantom{+}0.2541\pm 0.0161$& \nodata& $ \phantom{+}3.193\pm 0.589$& $ \phantom{+}1.727\pm 0.426$\\
\multicolumn{2}{l}{A2734} & 0.080 & $ \phantom{+}1.237\pm 0.344$& \nodata& \nodata& \nodata& \nodata& \nodata& \nodata\\
\tableline
\multicolumn{3}{l}{{\bf NFPJ001041.8-283444}} & $ \phantom{+}0.079\pm 1.049$& $ -2.570\pm 1.570$& $ \phantom{+}0.0279\pm 0.0415$& $ \phantom{+}0.0929\pm 0.0489$& $ \phantom{+}2.060\pm 0.657$& $ \phantom{+}5.373\pm 1.063$& $ -2.085\pm 0.722$\\
$-0.08$ & $\phantom{+}0.01$ & $\phantom{+}0.21$ & $ -5.704\pm 1.222$& $ \phantom{+}2.755\pm 1.470$& $ \phantom{+}0.996\pm 0.747$& $ \phantom{+}4.164\pm 0.983$& $ \phantom{+}3.711\pm 1.406$& $ \phantom{+}2.386\pm 0.532$& $ \phantom{+}2.425\pm 0.356$\\
18190 & \multicolumn{2}{c}{$1.988\pm0.053$}& $ \phantom{+}5.083\pm 1.145$& $ \phantom{+}2.890\pm 0.536$& $ \phantom{+}0.0154\pm 0.0136$& $ \phantom{+}0.1604\pm 0.0173$& \nodata& $ \phantom{+}2.252\pm 0.637$& $ \phantom{+}1.890\pm 0.458$\\
\multicolumn{2}{l}{A2734} & 0.080 & $ \phantom{+}0.539\pm 0.383$& \nodata& \nodata& \nodata& \nodata& \nodata& \nodata\\
\tableline
\multicolumn{3}{l}{{\bf NFPJ001042.6-284916}} & $ \phantom{+}0.077\pm 0.758$& $ -1.731\pm 1.139$& $ \phantom{+}0.0586\pm 0.0329$& $ \phantom{+}0.0961\pm 0.0382$& $ \phantom{+}1.571\pm 0.514$& $ \phantom{+}4.958\pm 0.858$& $ -0.832\pm 0.546$\\
$-0.60$ & $-0.80$ & $\phantom{+}0.20$ & $ -3.647\pm 0.947$& $ \phantom{+}4.662\pm 1.085$& $ \phantom{+}1.342\pm 0.577$& $ \phantom{+}2.329\pm 0.890$& $ \phantom{+}6.974\pm 1.150$& $ \phantom{+}1.751\pm 0.460$& $ \phantom{+}1.754\pm 0.304$\\
18682 & \multicolumn{2}{c}{$1.942\pm0.047$}& $ \phantom{+}4.843\pm 0.982$& $ \phantom{+}3.684\pm 0.469$& $ \phantom{+}0.0529\pm 0.0125$& $ \phantom{+}0.2069\pm 0.0163$& \nodata& $ \phantom{+}1.864\pm 0.580$& $ \phantom{+}2.101\pm 0.441$\\
\multicolumn{2}{l}{A2734} & 0.080 & $ \phantom{+}0.869\pm 0.340$& \nodata& \nodata& \nodata& \nodata& \nodata& \nodata\\
\tableline
\tableline
\end{tabular}
Table~\ref{tab:fullindexdata} is presented in its entirety in the electronic 
edition of the Astrophysical Journal. A portion is shown here for guidance 
regarding its form and content.
\end{table*}

The final galaxy sample will be culled for the analysis of the
linestrength-$\sigma$ trends and age and metallicity scaling relations
in Sections~\ref{sec:ixsig} and \ref{sec:trends}.

\section{Linestrength--$\sigma$ relations}
\label{sec:ixsig}

The overall trends of linestrengths with velocity dispersion reflect
changes in the characteristic stellar populations as a function of
galaxy mass.  In this section we describe the linestrength-$\sigma$
relations observed in the NFPS data and compare, where possible, to
previous results.

\subsection{Sample}

In the remainder of this paper, we will analyze a restricted sample of
the NFPS galaxies, hereafter referred to as the {\sc CULL}
dataset. This is defined by applying the following cuts:
\begin{enumerate}
\item
Galaxies must belong to one of the clusters in the NFPS cluster
sample, as determined in NFPS-I. This rejects objects in the field and
in background groups or clusters.
\item 
Galaxies must be free of emission lines as determined in this paper,
having emission weaker than 0.8\,\AA\ in OIII5007 {\it and} weaker
than 0.6\,\AA\ in \hb.
\end{enumerate}
Moreover, in the {\sc CULL} dataset, the velocity dispersions and
linestrengths are corrected for aperture effects as described in
Section~\ref{subsec:corrs}.

\subsection{Analysis of Individual Linestrengths with Velocity Dispersion}

\begin{figure*}
\center{
\includegraphics[height=140mm,width=170mm,angle=0]{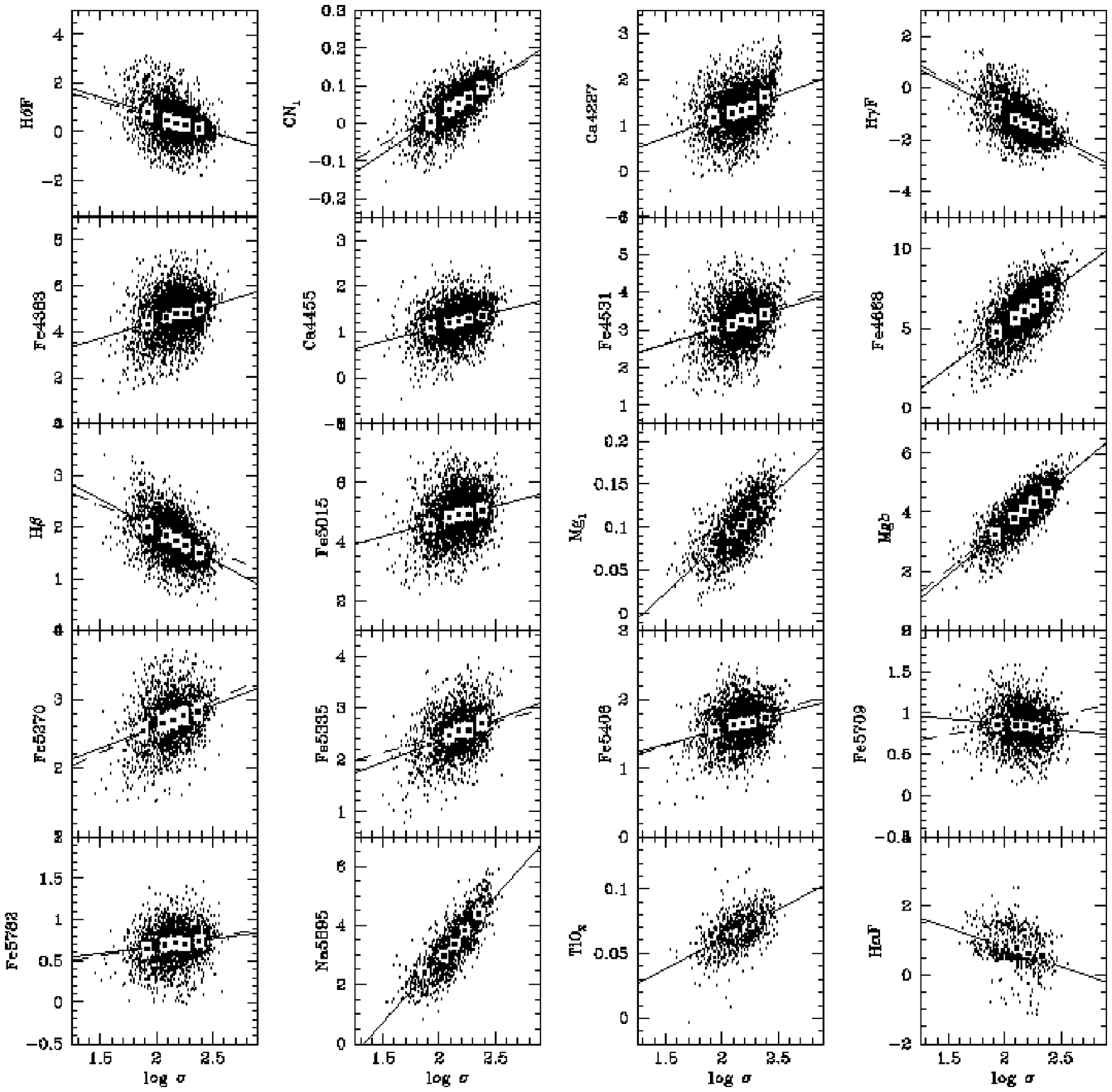}
}
\vskip -5mm
\caption{Our linestrength-$\sigma$ relations for each index.  The linestrength 
and velocity dispersion for the galaxies (points) are sorted into 
five bins by increasing velocity dispersion (black boxes outlined in 
white).  Solid lines are
the slopes from the linear regression of linestrength on log$\sigma$.
Dashed lines represent slopes predicted from our derived trends of
\afe, \zh, and age with velocity dispersion in
Section~\ref{sec:trends} (see text in that section).}
\label{fig:ixsig}
\end{figure*}

We restrict our attention to the 20 linestrength indices which are
least affected by systematic errors.  In fitting our
linestrength-$\sigma$ data, data points more than $4\sigma$ away from
the initial weighted least-squares fit are excluded. We then perform a
linear regression of linestrength on log$\sigma$ on the remaining
data, allowing for measurement errors in both variables but assuming
that any intrinsic scatter comes solely from the linestrengths.
Table~\ref{tab:ixtab} summarizes the parameters of our fits.  Slopes
and errors are quoted at both full (instrumental) and Lick resolution.
In Figure~\ref{fig:ixsig}, the linestrength-$\sigma$ relations for 20
indices are shown, both for individual data points and for means in
velocity dispersion bins.  The solid curves show the fitted slopes.

%\begin{deluxetable*}{ccccc}
\begin{table*}
%\tablecaption{Linestrength-$\sigma$ Fits}
\caption{Linestrength-$\sigma$ Fits}
\label{tab:ixtab}
\center
\begin{tabular}{ccccc}
\tableline
\tableline
Linestrength &
Units &
$N_{Galaxies}$\tablenotemark{a} &
Mean Slope (Lick)\tablenotemark{a} &
Mean Slope (Inst)\tablenotemark{b}
\\
\tableline
%\tablewidth{0pt}
%\tablehead{
%\colhead{Linestrength}   &
%\colhead{Units}   &
%\colhead{$N_{Galaxies}$\tablenotemark{a}}   &
%\colhead{Mean Slope (Lick)\tablenotemark{a}}   &
%\colhead{Mean Slope (Inst)\tablenotemark{b}}   }
%\startdata
 H$\delta_{\rm A}$ & \AA & 3419 & -3.484 $\pm$ 0.109 & -3.381 $\pm$ 0.116  \\ 
 H$\delta_{\rm F}$ & \AA & 3399 & -1.444 $\pm$ 0.055 & -1.408 $\pm$ 0.065  \\ 
 CN$_{1}$      & mag & 3435 & 0.197 $\pm$ 0.003  &  0.193 $\pm$ 0.004  \\ 
 Ca4227        & \AA & 3420 & 0.912 $\pm$ 0.039  &  0.421 $\pm$ 0.042  \\ 
 H$\gamma_{\rm A}$ & \AA & 3428 & -3.427 $\pm$ 0.096 & -2.997 $\pm$ 0.099  \\ 
 H$\gamma_{\rm F}$ & \AA & 3434 & -2.163 $\pm$ 0.052 & -2.222 $\pm$ 0.056  \\ 
 Fe4383        & \AA & 3437 & 1.448 $\pm$ 0.084  &  0.990 $\pm$ 0.088  \\ 
 Ca4455        & \AA & 3457 & 0.631 $\pm$ 0.033  &  0.451 $\pm$ 0.040  \\ 
 Fe4531        & \AA & 3442 & 0.917 $\pm$ 0.048  &  0.643 $\pm$ 0.056  \\ 
 Fe4668        & \AA & 3443 & 5.230 $\pm$ 0.109  &  4.885 $\pm$ 0.112  \\ 
 H$\beta$      & \AA & 3450 & -1.171 $\pm$ 0.032 & -1.166 $\pm$ 0.037  \\ 
 Fe5015        & \AA & 3450 & 1.029 $\pm$ 0.068  &  1.008 $\pm$ 0.077  \\ 
 Mg$_{1}$      & mag & 2018 & 0.121 $\pm$ 0.003  &  0.123 $\pm$ 0.002  \\ 
 Mg$_{2}$      & mag & 1993 & 0.189 $\pm$ 0.003  &  0.184 $\pm$ 0.003  \\ 
 Mg{\it b}     & \AA & 3414 & 3.201 $\pm$ 0.041  &  3.047 $\pm$ 0.043  \\ 
 Fe5270        & \AA & 1954 & 0.620 $\pm$ 0.041  &  0.614 $\pm$ 0.045  \\ 
 Fe5335        & \AA & 2113 & 0.821 $\pm$ 0.045  &  0.854 $\pm$ 0.053  \\ 
 Fe5406        & \AA & 2825 & 0.432 $\pm$ 0.029  &  0.432 $\pm$ 0.033  \\ 
 Fe5709        & \AA & 2503 & -0.131 $\pm$ 0.022 & -0.096 $\pm$ 0.026  \\ 
 Fe5782        & \AA & 2382 & 0.172 $\pm$ 0.024  &  0.289 $\pm$ 0.027  \\ 
 Na5895        & \AA & 1514 & 4.276 $\pm$ 0.075  &  4.590 $\pm$ 0.084  \\ 
 TiO$_{1}$     & mag &  789 & 0.021 $\pm$ 0.002  &  0.021 $\pm$ 0.002  \\ 
 TiO$_{2}$     & mag &  916 & 0.046 $\pm$ 0.002  &  0.046 $\pm$ 0.003  \\ 
 H$\alpha$A    & \AA &  542 & -0.722 $\pm$ 0.148 & -0.791 $\pm$ 0.067  \\ 
 H$\alpha$F    & \AA &  684 & -0.952 $\pm$ 0.104 & -1.147 $\pm$ 0.099  \\ 
%\enddata
\tableline
\tablenotetext{a}{
at Lick resolution.
}
\tablenotetext{b} {
at full (instrumental) resolution.
}
%\label{tab:ixtab}
\end{tabular}
\end{table*}
%\end{deluxetable*}

Of all the line indices, magnesium correlates most strongly with
$\sigma$ and has been most widely used as an indicator of stellar
populations.  In order to compare our Mg-$\sigma$ fit with those from
other surveys, we convert our Mg{\it b} line in \AA\ to Mg{\it b}$'$
in magnitudes using the conversion defined in \cite{Colless99}:
\begin{equation}
\mathrm{Mg}{\it b}' = -2.5 \log_{10}\left( 1 - \frac{\mathrm{Mg}{\it
b}}{32.5} \right).
\end{equation}
For $\sim$3400 galaxies, the Mg{\it b}$'$ linestrength is fit with the
relation $\Delta$ Mg{\it b}$' \propto$ (0.122$\pm$0.002) $\Delta (\log
\sigma)$.  This slope is very close to that found by EFAR,
(0.13$\pm$0.017), SDSS (0.15$\pm$0.02) and \cite{Kunt01}
(0.142$\pm$0.013).  When the data from \cite{Kunt01} are fit using our
fitting method, the slope (0.113) is closer to ours.  Our intrinsic
scatter of 0.011 mags is lower than both those of EFAR and
\cite{Kunt01}; we do not compare our scatter to that of SDSS since
they effectively reduce their scatter by creating composite spectra
from individual galaxies.

We take the logarithm of our \hb\ linestrengths to compare the
subsequent trends with $\sigma$ with other surveys: our \hb\ fit
yields log H$\beta \propto$ ($-0.306 \pm 0.008$) log $\sigma$ for 3440
galaxies compared to a slope of $-0.24\pm0.03$ for SDSS.
\cite{Jorg97} found log \hb\ $\propto$ ($-0.23 \pm 0.08$) log $\sigma$
with an intrinsic scatter of 0.061 compared to our intrinsic scatter
of 0.041.  We discuss the effects on our conclusions of a different \hb\ 
slope in Section~\ref{sec:trenderrs}.

In order to compare our iron relations with other surveys, we defined
$<$Fe$>$ as
\begin{equation}
\mathrm{<Fe>} = \frac{\mathrm{Fe}5270 + \mathrm{Fe}5335}{2}.
\end{equation}
For 1090 galaxies with both Fe5270 and Fe5335 measurements, our best
fit to log $\sigma$ is log $<$Fe$> \propto$ (0.088$\pm$0.008)
log$\sigma$.  This is consistent with the lower value found by
\cite{Jorg97} (0.075$\pm$0.025) and the higher value in the SDSS
(0.11$\pm$0.03).

The linestrength Fe5709 has a negative slope against $\log \sigma$,
and is the only iron line to behave in this manner.  While it is
possible that the line would be contaminated by strong sky emission
from the Na I$\lambda\lambda$5683,5688 lines, this would only affect
measurements in the extremely low-redshift clusters.  We have analyzed
the linestrength data from \cite{Trager98}, and, for Fe5709, find a
slope of -0.06$\pm$0.09, which is consistent with our result.

We have shown that each of the 20 linestrengths considered shows
significant dependence on velocity dispersion.  The pattern which
emerges is that all Balmer lines decrease with increasing velocity
dispersion, whereas all other lines (with the exception of Fe5709)
increase with increasing velocity dispersion.  The following section
describes a robust method for interpreting these trends as scaling
relations of stellar population properties, as a function of galaxy
mass.

\section{Global Age and Metallicity Trends}
\label{sec:trends}

Stellar population models, such as those by \cite{TMB03} (hereafter
TMB) and the extension of these which includes H$\gamma$ and H$\delta$
\citep{TMK04}, predict linestrengths from a range of stellar
population parameters.  These models are an extension of those by
\cite{Maraston98} with adjustments based on theoretical stellar
atmosphere calculations.  TMB are currently the only models that
predict linestrengths for populations with non-solar abundance ratios,
as required for elliptical galaxy studies.  TMB do not, however,
include possible contributions from blue Horizontal Branch (BHB)
stars, which might arise from a low-metallicity subpopulation
\citep{Maraston00}, or from enhanced mass-loss in evolved stars
\citep{Thomas04}.  Whatever their origin, such stars would contribute
to a strengthening of the Balmer absorption lines, mimicking the
effect of younger ages. In principle, BHB stars have greater impact on
the high-order lines than on \hb\ which could ultimately provide a
means to distinguish their effects from those of a younger population
\citep{Schiavon04}.  In this paper, we will assume that the BHB
contribution is either negligible, or at least not strongly dependent
on galaxy mass.

\subsection{Method}
\label{sec:method}

We wish to use our linestrength data to determine ages, metallicities,
and $\alpha$-element enhancements as a function of galaxy mass. There
are several ways to extract this information. The usual method,
adopted by most previous studies, is to derive these parameters for
individual galaxies by interpolating (sometimes extrapolating) the
model grids.  However, when the measurement errors in the
linestrengths are non-negligible (as is the case with the NFPS data),
the tilt of the model grid leads to correlated errors in age and
metallicity for each galaxy \citep{Kunt01}.  These correlated errors
complicate the interpretation of the data, potentially generating a
spurious age-metallicity correlation \citep{Terlevich02}.

This problem can be overcome by stacking the spectra of similar
galaxies, for example galaxies within a narrow bin of velocity
dispersion \citep{Bernardi03}, to create a composite spectrum of high
signal-to-noise ratio. Equivalently, one can average the linestrength
measurements themselves for galaxies in each bin.  In either case, one
averages over scatter in the linestrengths that may be due to
measurement errors, but also averages the scatter due to intrinsic
(possibly correlated) dispersion in the galaxy population
parameters. The obvious weakness of this approach is that one can only
state confidently the characteristics of the mean galaxy; thus, there
is no information regarding the variation in parameters from galaxy to
galaxy within the bin.

\begin{figure*}
\includegraphics[width=140mm,angle=270]{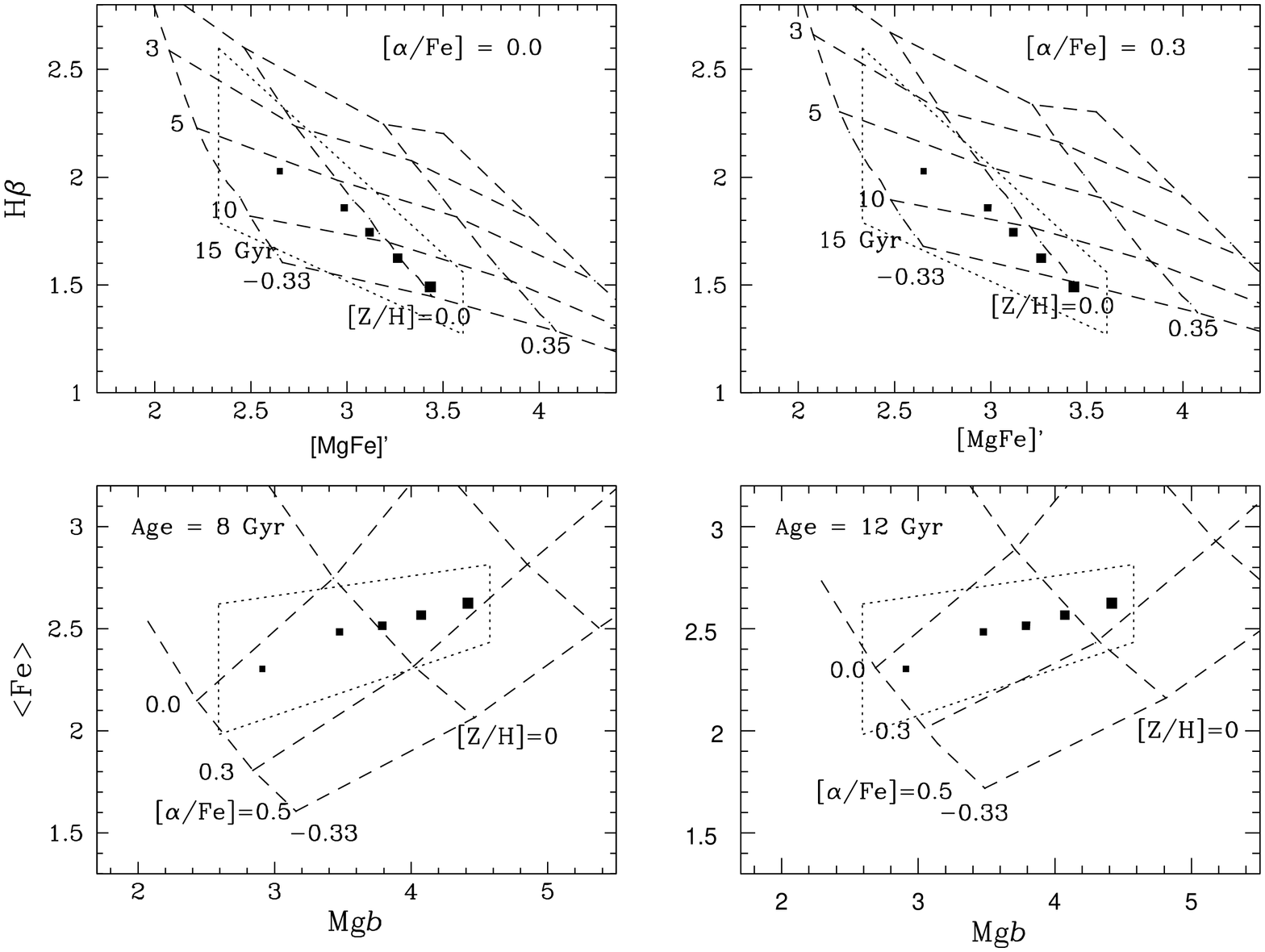}
\vskip -2mm
\caption{Model grids from stellar population models by \cite{TMB03}.  Each data
point is the mean of a range of velocity dispersion, with each bin
having approximately the same number of galaxies.  The size of the
points increases as $\sigma$ increases.  The area inside the dotted
line is the range covered by the average errors of the linestrengths.
Linestrengths are in units of \AA.}
\label{fig:grids}
\end{figure*}

As a preliminary step, we plot our binned linestrengths on top of
grids derived from linestrength-parameter scaling relations from TMB
in Figure~\ref{fig:grids}.  The quantity [MgFe]$^\prime$ in the lower
two panels was defined by \cite{TMB03} as
\begin{equation}
\mathrm{[MgFe]^\prime \equiv \sqrt{Mg{\it b}(0.72 \times Fe5270+0.28
\times Fe5335)}.}
\label{eqn:mgfep}
\end{equation}
This index was defined to be sensitive to overall metallicity, but
almost independent of \afe.  In Figure~\ref{fig:grids}, the area
inside the dotted trapezoid represents the average error for a {\it
single} linestrength measurement in each bin.  Since each bin
represents $\sim$700 galaxies, the errors in the mean values are very
small, and not plotted here.

In the upper-left panel of Figure~\ref{fig:grids} where we plot models
with \afe\ = 0.0, our five points form a line of almost constant
metallicity \zh\ = 0.0 along with an increasing age estimate as the
velocity dispersion increases.  In the upper-right panel where \afe\ =
0.3, we can see a slight shift of our data points toward higher
metallicity and again toward older ages, with increasing $\sigma$.  In
the lower panels, the grids show predictions for constant age, while
our five binned points follow a rough trend in metallicity --
increasing \zh\ with increasing velocity dispersion.  There is a
slightly less pronounced trend of increasing \afe\ with increasing
$\sigma$. It should be realized that, since each panel shows a
two-parameter projection of the three-parameter models, it is not
trivial to read off the ``correct'' values from these plots.

The stacking methods described above work well for a triplet of
linestrengths (e.g.  \hb, Mg{\it b}, $<$Fe$>$), from which one derives
the triplet of galaxy parameters age, metallicity and
alpha-enhancement.  However, in general, each choice of linestrength
triplet will lead to different stellar population parameters.
Ideally, we seek a method to combine the information from all
linestrengths simultaneously.  An obvious approach is to fit, via a
$\chi^2$ method, all linestrengths as a function of the galaxy
parameters \citep{Proctorchi2}.  Unfortunately, the zero-points are
typically uncertain, both for the model predictions and for the
observational measurements. As a result, the model grids often
systematically under- or over-predict certain linestrengths. Inverting
the grids then yields estimated ages (and other parameters) that are
systematically in error. In a $\chi^2$ approach, the linestrengths
which are affected in this way will be outliers in the fit and will
carry disproportionate weight in the results.

Here, we choose instead to describe the whole sample simultaneously,
in terms of a set of scaling relations, with log(age), [Z/H] and \afe\
each following a linear trend with $\log\sigma$.  Here [Z/H] is the
total metallicity, expressed logarithmically relative to solar, while
\afe\ is the $\alpha$-element enhancement, similarly expressed.  These
global trends can be estimated by modeling their combined effects on
the observed linestrength-$\sigma$ relation.  We can find the
parameter-$\sigma$ relations from
\begin{equation}
\frac{dI}{d\log \sigma} = \sum_{i=1}^N{\frac{dI}{dP}\frac{dP}{d\log \sigma}}
\label{eqn:sum}
\end{equation}
where $I$ is the linestrength (at Lick resolution), $P$ is a parameter
from the stellar population models (log age, [Z/H] or \afe),
d$I$/d$\log \sigma$ are our linestrength-$\sigma$ slopes from the
previous section, and d$I$/d$P$ are the linestrength-parameter
``responses'' from the TMB models.  While the absolute zero-point of
the grids is uncertain, the d$I$/d$P$ responses are more robust.  We
emphasize that our d$I$/d$\sigma$ relations involve the averaging of
$\sim$3000 linestrength measurements for each index; thus, statistical
errors are suppressed and do not significantly influence the results
of the regression.  The aim of this analysis is to constrain the
d$P$/d$\log \sigma$ representing the slope of the parameter scaling
relations.

This differential method has a number of advantages over the ``grid
inversion'' approach.  As already mentioned, it avoids the problem of
interpreting correlated errors in the derived parameters. Also it uses
only the {\it relative} changes in the predicted linestrengths from
the TMB models and the {\it relative} changes in our measured
linestrengths with velocity dispersion.  Our method is thus explicitly
insensitive to calibration uncertainties in the models and to overall
zero-point errors in the linestrength measurements.  Finally, unlike
stacking spectra or linestrength measurements by velocity dispersion,
there is no need to bin the data into arbitrarily-defined subsets.
The main shortcomings of the method are its assumptions that the model
grids are parallel over the parameter-space spanned by our sample and
that the parameter-$\log\sigma$ relations are indeed linear.

The outputs of the differential method are the parameter-$\sigma$
scaling relations, which depend on the choice of the input model
responses d$I$/d$P$.  To estimate d$I$/d$P$, we calculate $\Delta
I/\Delta P$ from the model grids themselves by differencing an upper
(or lower) value with a fiducial value. For example, our data span
3--15 Gyr in age, with a central age of 8 Gyr.  Referring to
Fig~\ref{fig:grids}, for a metallicity of \zh\ = 0.0 and \afe=0.0, we
see that $\Delta$H$\beta_{\rm pred} = -0.288$ between 5 and 10 Gyr.
Note that the grids are very nearly parallel.  Thus, if we had assumed 
\zh\ = 0.35 instead, we would have obtained a very similar value, 
$\Delta$H$\beta_{\rm pred} = -0.294$.  To obtain the best estimates of the
model responses, we need to make reasonable choices for the central
values of the three parameters, and for their upper and lower
values. We adopt as a central value the parameters from our fits
corresponding to the median velocity dispersion (log $\sigma \sim$
2.1): \afe\ = 0.2, age = 8 Gyr, and \zh\ the average of 0.0 and
0.35. The range for each parameter was restricted to the range spanned
by the sample galaxies, i.e. age: 3--15 Gyr, \afe: 0.0--0.5, and \zh:
--0.33--0.67.  Because the grids are not exactly parallel, there is a
small uncertainty in our derived parameter scaling relations
introduced by our choice of central values.  We will discuss this in more
detail in Section~\ref{sec:trenderrs}

To fit three model parameters, we need at least
three linestrength-$\sigma$ relations.  We can over-constrain the fit
by including more than three indices, and in practice we use the
following 12 line indices: CN$_{1}$, \hb, H{$\gamma_{\rm F}$,
H{$\delta_{\rm F}$, Fe4531, Fe4668, Mg{\it b}, Fe5270, Fe5335, Fe5406,
Fe5709, and Fe5782.  This set includes three age-sensitive Balmer
lines, two indices quite sensitive to $\alpha$-enhancement (Mg{\it b},
CN$_1$) and a range of mainly metallicity-sensitive features.  We did
not include lines such as Fe4383 and Fe5015 where the systematic
correction to the linestrength is a significant factor of the random
error (see Section~\ref{subsec:errors}).  Ca4227 was excluded because
in general the Ca abundances are not very well understood
\citep{Saglia02}, and we avoid including partially redundant indices,
for instance H$\gamma_{\rm A}$.  Alternative choices of indices are
possible and the selection of which to include can affect the fit
results by more than the formal random errors. We discuss this issue
further in Section~\ref{sec:trenderrs}.

The d$I$/d$\log \sigma$ slopes we use from our data derive from the
regression of linestrength on log$\sigma$, as in
Table~\ref{tab:ixtab}. To weight contributions from the different line
indices in the fit, we use the formal regression errors on the
linestrength$-\sigma$ slopes. No contribution from model error is
included.

\subsection{Results}
  
When all 12 line indices are used in the fit, we obtain
\dafe$^{0.31}$, \dzh$^{0.53}$ and \dage$^{0.59}$.  The formal error
estimates are very small (0.01, 0.02 and 0.02, respectively), and, as
we discuss in greater detail below, are not representative of the true
uncertainties.  For more insight into which linestrengths drive the
fit results, we have explored restricted models in which one or more
of the parameters is held constant with $\sigma$.  In
Figure~\ref{fig:slopevsslope} we show the predicted versus observed
linestrength-$\sigma$ slopes for the 12 line indices when different
combinations of the stellar parameters are allowed to vary with
$\sigma$.  We normalize both the ``actual slope'' (i.e., our measured
linestrength-$\sigma$ slope) and the predicted slope by dividing by
the error in our linestrength-$\sigma$ slopes.  Thus, the diagonal
line in each panel shows where the actual and predicted slopes agree.
Any vertical displacement of a linestrength indicates an under- or
over-prediction of the observed slope.

\begin{figure*}
\center{
\includegraphics[height=180mm,angle=0]{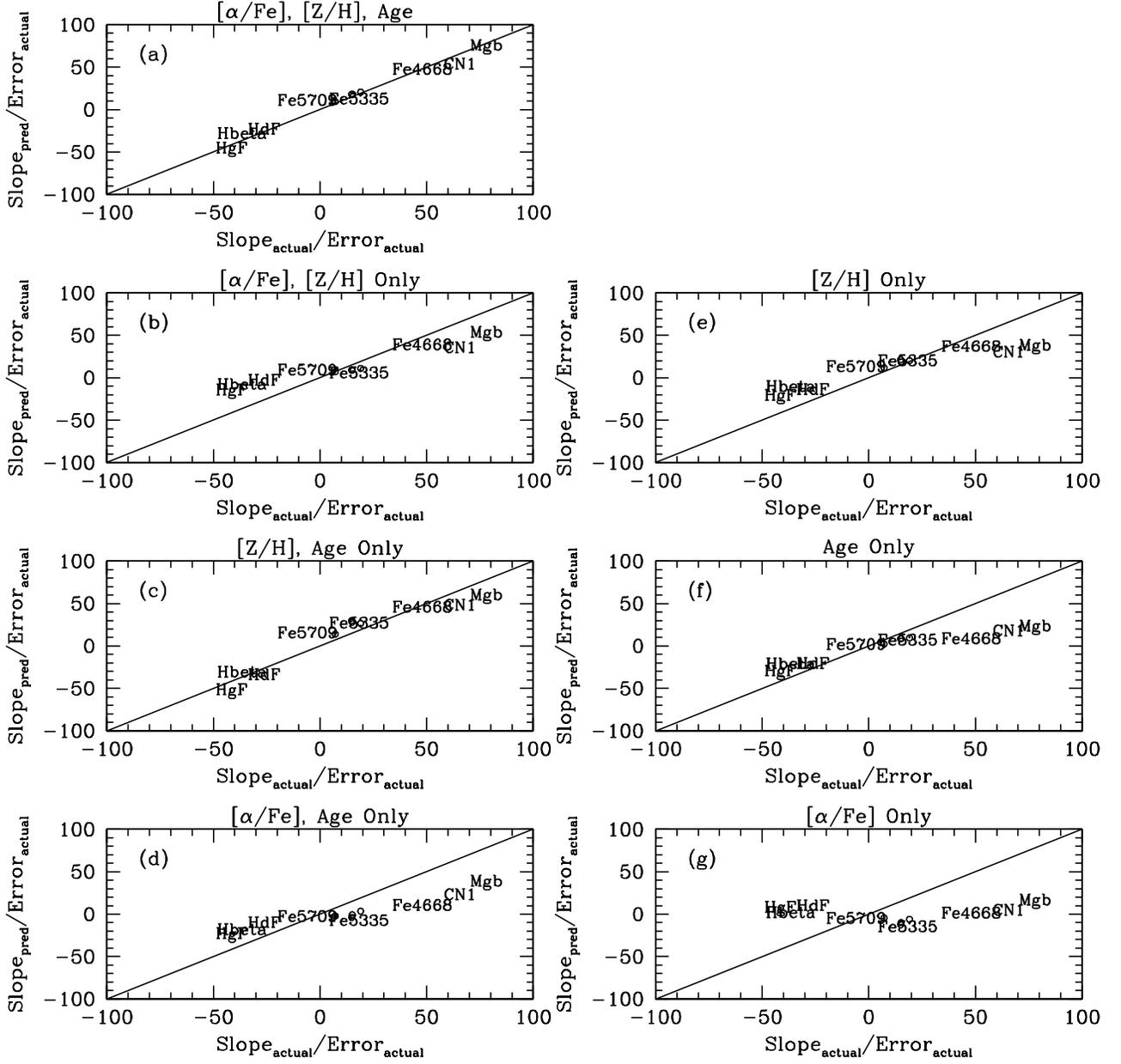}
}
\vskip -1mm
\caption{A comparison of the predicted index-$\sigma$ slopes (vertical axis)
with the observed slopes (horizontal axis) for different combinations
of the three stellar population parameters.  Labels indicate the
results for each of the 12 linestrength indices. The diagonal line
indicates perfect agreement between model-predicted and measured
slopes.  Note that both model-predicted slopes and measured slopes are
normalized by the formal error in the measured slope. Thus the
vertical offset from the line indicates the degree to which the model
fails to predict the observed slope. Panel (a) shows our default
solution in which \afe, \zh\ and log(age) are scaled with
log$\sigma$.  The model predictions for Fe5709 deviate from the
observed slope at a high significance level (compare with Fig
8). Note, however that Fe5709 is discrepant in all panels. Panels
(b)-(d) show the comparisons when we assume no trend with $\sigma$ for
age, \afe\ and \zh\ respectively. In these cases, there are multiple
indices which deviate strongly from the model predictions. For
example, if we assume no age gradient (panel b), we cannot
simultaneously reproduce the slopes of Mg{\it b} and CN$_{1}$ and the Balmer
lines. The fits are considerably poorer when only one parameter is
allowed to vary, as in panels (e)-(g).}
\label{fig:slopevsslope}
\end{figure*} 

Figure 10a shows good agreement between the actual and predicted
linestrength-$\sigma$ slopes when \afe, \zh, and age all vary with
$\sigma$, i.e. our default solution.  When age is forced to be
constant with $\sigma$, as in Figure 10b, the predicted (negative)
slopes of the Balmer lines are too shallow and the predicted
(positive) slopes of the $\alpha$-sensitive indices, Mg{\it b} and
CN$_{1}$, are also too low.  Thus, without an age--$\sigma$ relation,
we cannot simultaneously reproduce both the Balmer lines and the
$\alpha$-sensitive slopes.

We also see that if metallicity is held constant with $\sigma$, as in
Figure 10d, the predicted slopes of the Balmer lines are too steep, as
is that of Mg{\it b}.  Similar arguments apply to the other restricted
fits.  In the case where only \afe\ is allowed to vary with $\sigma$
(Figure 10g), the fit fails catastrophically, since a
correctly-predicted Mg{\it b} slope would require \afe\ increasing
with $\sigma$, while a decreasing \afe\ with $\sigma$ would be needed
to produce the observed negative Balmer line slopes. The result is a 
very poor fit with no \afe\ trend at all.

Using the derived parameter-$\sigma$ relations, we can re-derive our
linestrength-$\sigma$ fits and check the consistency with our
data-derived linestrength-$\sigma$ slopes.  We used our default
solution for the parameter-$\sigma$ relations (\dafe$^{0.31}$,
\dzh$^{0.53}$, and \dage$^{0.59}$) to generate the predicted slopes
(dashed lines) in Figure~\ref{fig:ixsig}.  For each linestrength, the
predicted slope is very similar to the measured slope with the
exception of Fe5709, whose measured linestrength-$\sigma$ slope is
opposite to those of the other iron line indices.

In Section~\ref{sec:ixsig}, we noted that the \hb\ slopes found by
both SDSS and \cite{Jorg97} are 20\% lower than ours. We have repeated 
the analysis using their slope but found no significant change in our
results, including the age--mass gradient.

\subsection{Error Estimation}\label{sec:trenderrs}

The formal errors on the galaxy parameter slopes d$P$/d(log$\sigma$)
are very small: specifically, we find \dafe$^{0.31\pm0.01}$,
\dzh$^{0.53\pm0.02}$, \dage$^{0.59\pm0.02}$.  From a purely
statistical point of view, these errors may be slightly 
underestimated for two reasons: first, the linestrength measurements
are not strictly independent since in a few cases the bandpasses
overlap (e.g., Fe5270 and Fe5335), and second, it is possible that the
scatter in intrinsic parameters from galaxy-to-galaxy may be
correlated (e.g., Trager et al. 2000b suggests that at a given
$\sigma$, younger galaxies are also more metal-rich).  Neither of
these effects will bias the resulting age, metallicity and
alpha-enhancement trends, which are essentially the means of the
galaxy parameters at a fixed sigma.  However, we will show below that 
systematic effects vastly dominate over the statistical errors,
so will use the former as our error estimate and neglect the latter. 
There are two sources of systematic error: one is the 
choice of linestrengths indices used in the regression; the other is 
due to the uncertainty in estimating the responses $dI/dP$. 

To estimate the systematic errors due to choice of linestrength indices, 
we divide our 12 line indices into three groups: Balmer lines (\hb,
H$\gamma_{\rm F}$, H$\delta_{\rm F}$), $\alpha$-sensitive indices
(Mg{\it b}, CN$_{1}$) and the seven Fe-dominated indices.  
We then examine the scaling relations that result when only one index 
from a given group is used, and all indices from the other two groups 
are used.  Using Figure~\ref{fig:ixefx} as a guide, we can draw several
conclusions.  First the only consistently ``outlying'' line indices 
are Fe5335 and Fe5782, which prefer a weaker age trend, a stronger
metallicity trend, and a weaker \afe\ trend, and Fe5709, for
which the reverse is true.  In general, there is little spread of the
scaling relations; choosing other combinations of linestrength indices yield
consistent results.  In particular, note that the age gradient is
essentially unchanged whichever Balmer line is used.  Note also that the
scatter along the age-metallicity degeneracy line is such that if the
age trend is steeper by 0.1 the metallicity trend is flatter by
$\sim$0.03.  We estimate the resulting errors on our scaling relations by 
calculating the standard deviation of the results from all of the linestrength 
combinations.  We find the deviations to be 0.03 in the 
\afe-log$\sigma$ relation, 0.06 in \zh-log$\sigma$, and 0.08 in 
age-log$\sigma$.

\begin{figure*}
\center{
\includegraphics[height=180mm,width=180mm,angle=0]{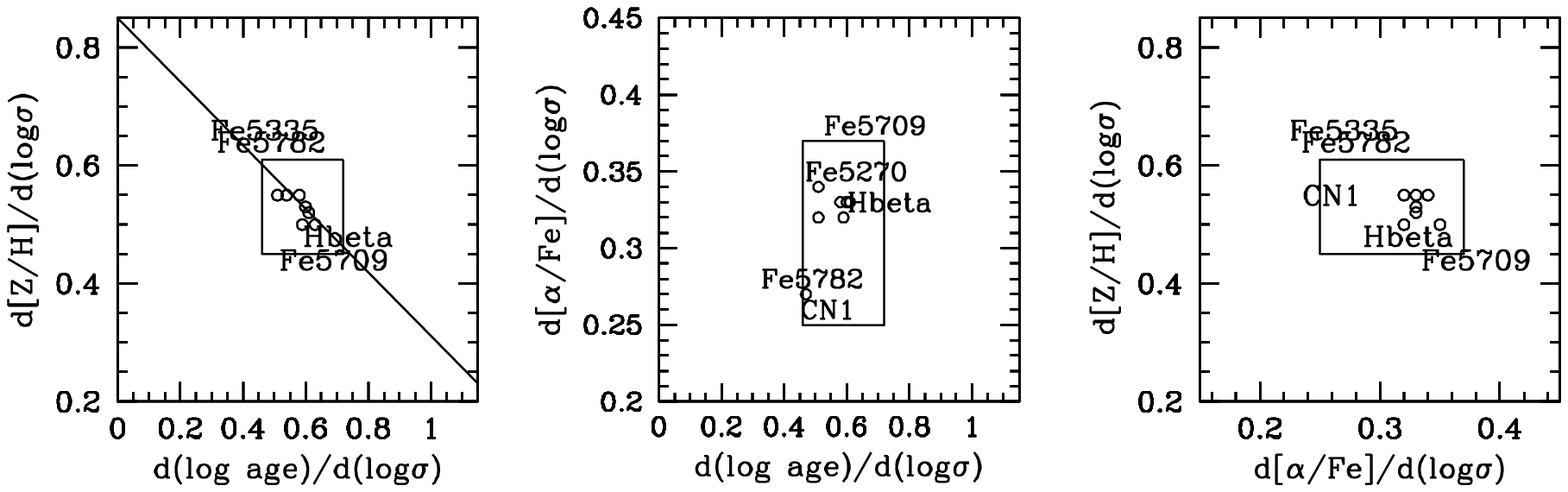}
}
\vskip -120mm
\caption{The effect of inclusion of certain linestrengths on our regression.
Each label or open circle 
represents the inclusion of that element and the exclusion
of the other elements {\it in that group} (see text).  
For example, the Fe5709 label indicates the fit
results when Fe5709 is the only Fe-dominated index included (all of
the Balmer and $\alpha$-sensitive indices are retained for this test,
however).  Similarly, the CN$_1$ label indicates that for that fit,
CN$_1$ was the {\it only} $\alpha-$index included (i.e., Mg$b$ was
excluded), but all of the Balmer and Fe indices were also included.
For clarity, only indices which have a significant effect on the default 
solution are labeled; other indices are shown as open circles. 
The straight line in the left panel is drawn by eye to represent the
linearity of the age-metallicity degeneracy.
The rectangle in each panel outlines our default solution with error range.}
\label{fig:ixefx}
\end{figure*} 

\begin{figure*}
\center{
\includegraphics[height=180mm,width=180mm,angle=0]{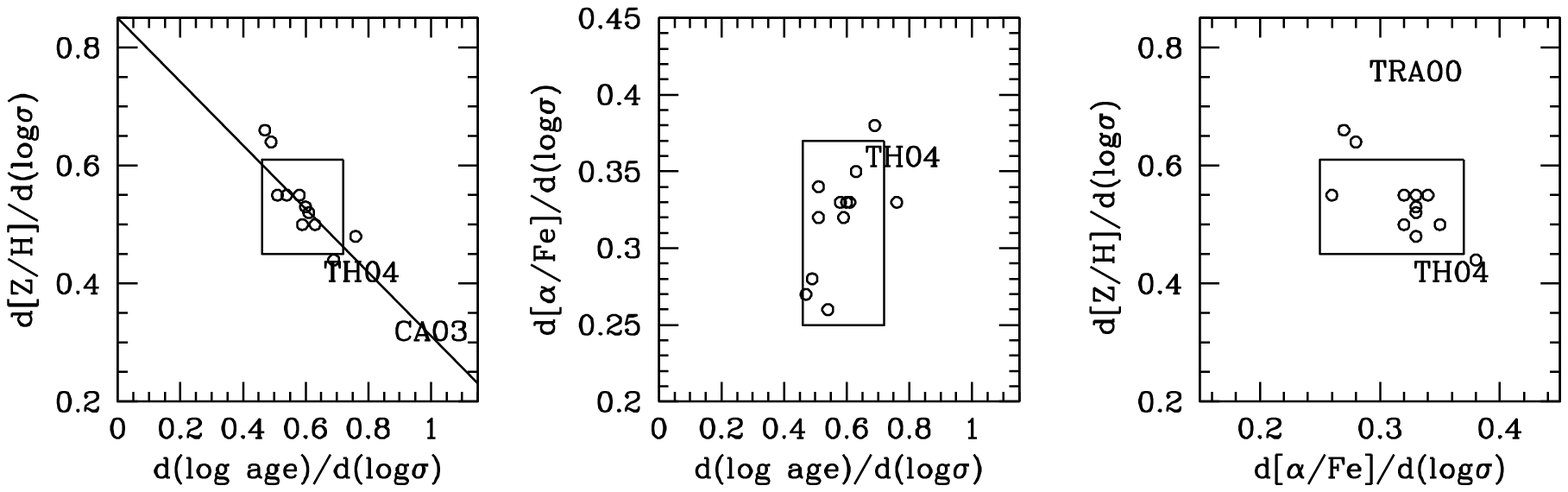}
}
\vskip -120mm
\caption{Our range of stellar population parameter
scaling relations with $\sigma$ plotted along with those from other
groups.  (CA03 = our estimate from Figure 21 and Table 9 from
\cite{Caldwell03}; TRA00 = \cite{Trager00b}; TH04 = \cite{Thomas04}).
Open circles represent the range in NFPS scaling relations when
certain linestrengths are included from the fit while the rest in
their group have been excluded (see text and Figure~\ref{fig:ixefx}).}
\label{fig:compare}
\end{figure*}

Our second error component is the deviation of our parameter-$\sigma$
scaling relations when different sets of d$I$/d$P$ values were used in
the regression.  We consider eight cases in which are central values of \afe, 
\zh, and age are perturbed by $^{+0.2}_{-0.1}$, $\pm0.35$ and $\pm2$ Gyr
respectively, and the responses d$I$/d$P$ accordingly recalculated
around these central values.  We find a standard deviation of 0.07 in
the \afe-log$\sigma$ slope, 0.08 in \zh-log$\sigma$, and 0.10 in
log(age)-log$\sigma$.  When these errors are added in quadrature with the
errors from different linestrength combinations, the total error is
0.06 in \afe-log$\sigma$, 0.08 in \zh-log$\sigma$, and 0.13 in
log(age)-log$\sigma$.

In summary, we see that the inclusion or exclusion of individual
indices and the choice of input d$I$/d$P$ values affect the results by
considerably more than the formal errors.  The scatter between various
results allows estimation of more realistic uncertainties.  Our final
results, with errors, are: \dafe$^{0.31 \pm 0.06}$, \dzh$^{0.53 \pm
0.08}$, and \dage$^{0.59 \pm 0.13}$.  These errors are correlated in
the sense that a stronger age trend corresponds to a weaker
metallicity trend and vice-versa.

\subsection{Grid inversion: a consistency check}

The slopes analysis of the previous section makes the assumption that
the grids are linear and parallel (i.e.\ that a single set of
responses d$I$/d$P$ is sufficient to capture the structure of the
models). Also, we imposed a linear relationship of each population
parameter with $\log\sigma$.  We have argued that despite these
restrictions, this explicitly differential method is more robust
against calibration uncertainties in both the data and the models,
than direct interpolation of model grids, and has the further
advantage that it does not require explicit binning.

In this section, we return to the grid interpolation method as a
consistency test, and to assess the limitations of the slopes
analysis. Since the individual measurements are subject to sizable
random errors, we use the linestrengths averaged over the five
velocity dispersion bins of Figure~\ref{fig:grids}.  The model grid is
inverted by determining, for each velocity dispersion bin, the
parameters (log age, [Z/H], \afe) which best reproduce (according to a
$\chi^2$ statistic) the same twelve line indices used above. This
analysis yields an estimate of the three population parameters as a
function of $\log\sigma$ which is not forced to the linear form
imposed in the slopes method. The results are shown in
Figure~\ref{fig:grinvert}.  The grid-inversion approach yields
age-$\sigma$, [Z/H]-$\sigma$ and \afe-$\sigma$ relations in excellent
agreement with results from our slopes method.  At face value, the age
of the most massive galaxies is $\sim$11\,Gyr, but for the reasons
emphasized elsewhere, the absolute values of age, [Z/H] and \afe\ are
less secure than the relative change along the mass sequence.  The
upper panel of Figure~\ref{fig:grinvert} suggests a non-linearity in
the age-log$\sigma$ relation, which appears to steepen by almost a factor
of two at the lowest masses (i.e., low $\sigma$). Such behavior is, by
definition, not observed in the slopes analysis of the previous
sections.

\begin{figure}
\includegraphics[width=170mm,angle=270]{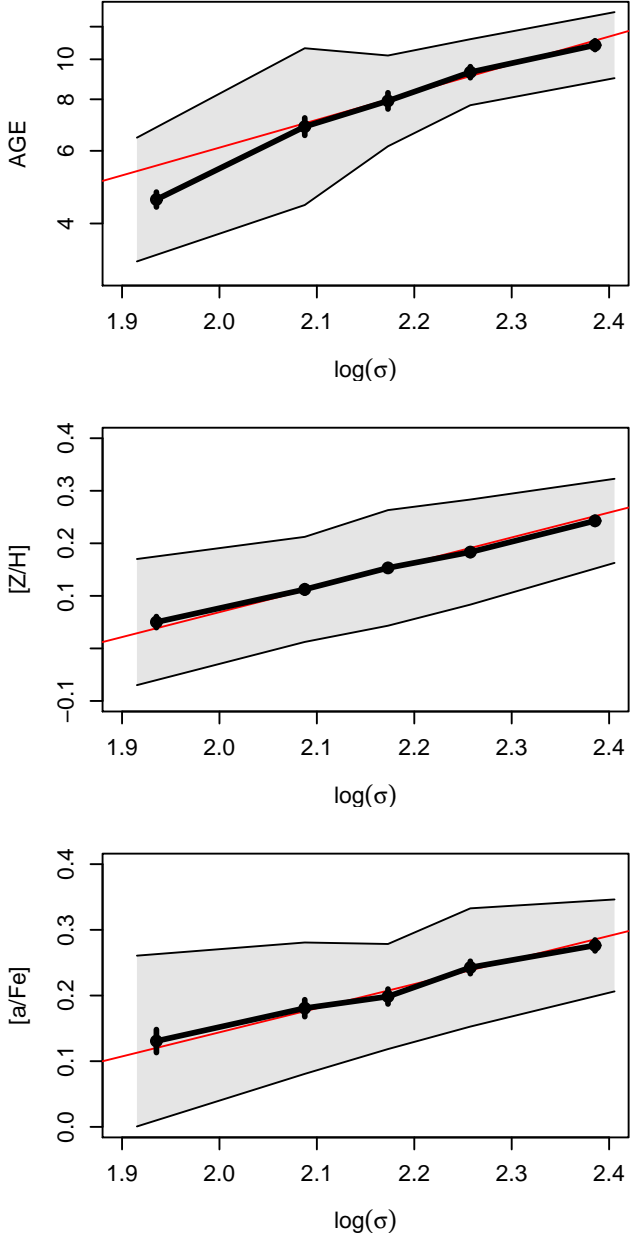}
\caption{Fundamental scaling relations of the population parameters with
$\sigma$.  The thick line shows average age, metallicity and \afe\
determined in bins of velocity dispersion, using the grid-inversion
method. The set of line indices employed is the same as that in the
slopes method. For comparison, the linear relations obtained from the
slopes method are overplotted as the thin red line, with zero-point
adjusted to match the grid-inversion results. The light-shaded regions
indicate the {\it maximum} internal scatter estimated in
Section~\ref{sec:internalscat}.}
\label{fig:grinvert}
\end{figure}

\subsection{Internal population scatter}\label{sec:internalscat}

An issue of considerable interest is the degree of internal scatter
among galaxy properties at a given point on the mass sequence. For
instance, the tightness of the color-magnitude relation has been used
to infer limits on the spread in formation ages of ellipticals
\citep{Bower92}.  The scatter around linestrength-$\sigma$ relations
in principle provides very powerful constraints, since each line index
has different sensitivities to the underlying population parameters.

While an upper limit to the scatter in linestrength-$\sigma$ relations
is readily available from the measured scatter, estimating the {\it
intrinsic} scatter requires accurate knowledge of the experimental
errors, and a robust estimation method.  In each bin, we first fit a
linear linestrength-$\sigma$ trend (allowing the slope and zero-point
to vary from bin to bin if the data require this). Then we model the
linestrength residuals assuming a constant intrinsic scatter 
for that bin $S(I)$,
in addition to the measurement errors (which include the systematic
components estimated in Section~\ref{subsec:errors}).  In this way we
can determine the range of $S(I)$ values which can reproduce the
observed inter-quartile range of the data. Relative to a standard
maximum likelihood estimate this is more robust against outliers, and
focuses on matching the core of the distribution.

The 95\% confidence limits on $S(I)$ are plotted in
Figure~\ref{fig:intrinscats} for a subset of our line indices, for the
five velocity dispersion bins used elsewhere in this paper. The line
indices shown are those in which intrinsic scatter is most cleanly
detected: Mg{\it b}, H$\gamma_{\rm F}$, Fe5406 and Fe4668.

\begin{figure}
\includegraphics[width=190mm,angle=270]{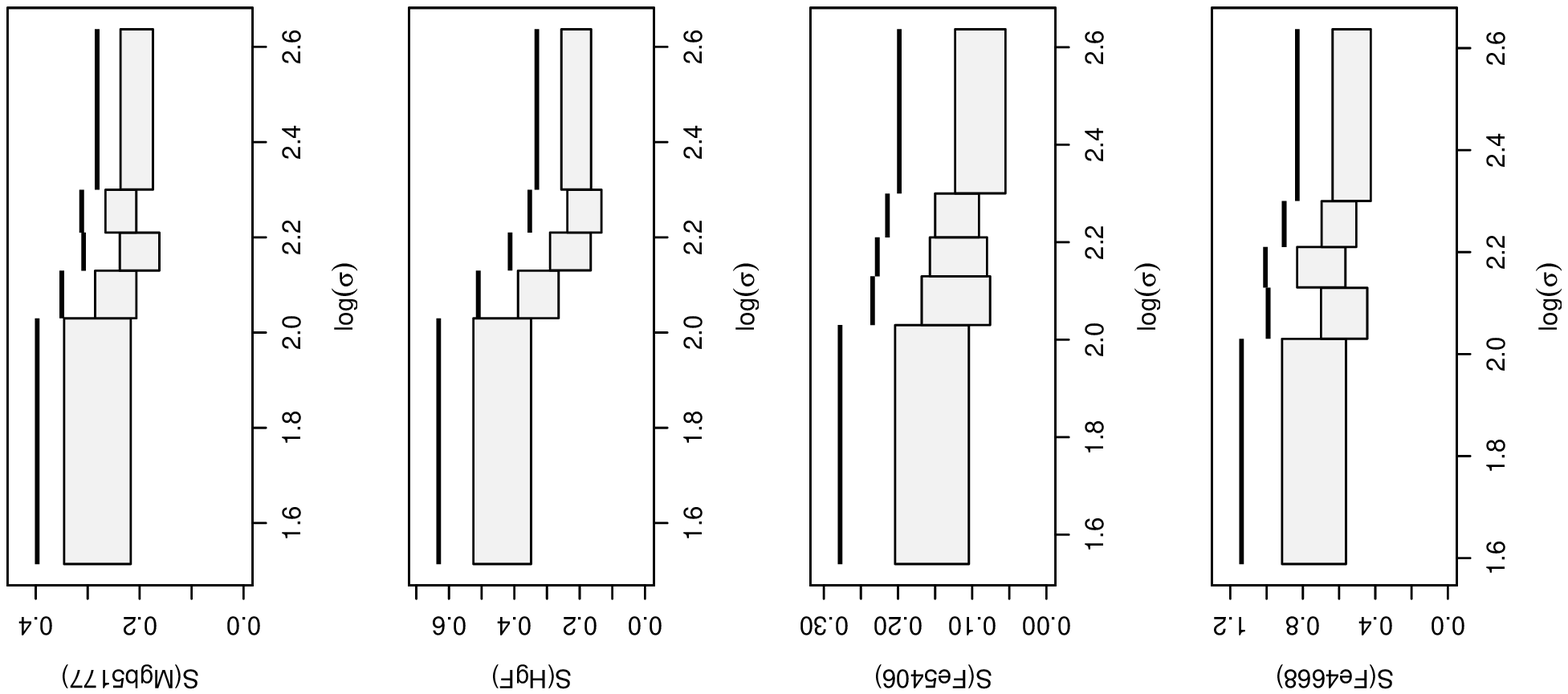}
\caption{Total, measurement and intrinsic scatter around the linestrength-$\sigma$ 
relationships. In each panel, the black lines show the observed (i.e., total) 
scatter in \AA\ (strictly, this is a robust estimate of the $1\sigma$ scatter, 
based on the measured interquartile range). The 95\% confidence intervals for 
the intrinsic $1\sigma$ scatter are indicated by the shaded boxes.}
\label{fig:intrinscats}
\end{figure}

Given a set of models such as TMB, the scatter $S(I)$ in each line
index yields an upper and lower bound on the internal age scatter
$S($log\ age$)$, and similarly for [Z/H] and \afe.  Using multiple
line indices, these constraints can be compounded, to leave a narrow
range in each parameter which is consistent with the observed
scatter. In this way, we have determined the internal dispersion
consistent with the observed scatters shown in
Figure~\ref{fig:intrinscats}.  To simplify the problem, we have
assumed that the parameter scatters are independent, i.e. within each
bin there is no internal correlation between age-$\sigma$ and
metallicity-$\sigma$ residuals.  In converting parameter scatter to
predicted linestrength scatters, we use model responses appropriate to
each bin, given the grid inversion results above.  Note that we do not
use \hb\ here; significant intrinsic scatter in \hb\ is observed, at a
level inconsistent with the other line indices under any model for the
underlying distribution. This is likely due to variations in low-level
residual nebular emission.

The results of these calculations are included in
Table~\ref{tab:paramfits}, along with the results of the grid
inversion analysis.  For each bin, we tabulate the range and average
of $\sigma$, the average stellar population parameters, and the range
of population scatter consistent with the limits in
Figure~\ref{fig:intrinscats}.  The parameter scatters should be used
with caution, since they are quite sensitive to the accuracy of our
observational error estimates.  Taken at face value, however, the
analysis suggests a $\sim$25\,\% scatter in metallicity at given
$\sigma$, while the age scatter appears to vary with mass, increasing
from $\sim$15\% at high $\sigma$ to $\sim$40\% for the low-$\sigma$
bin.

\begin{table*}
\caption{Age, metallicity and \afe\ by $\sigma$-bin}
\label{tab:paramfits}
\center
\begin{tabular}{lccccccc}
\tableline
\tableline
Range of $\sigma$ &              
$\langle\log\sigma\rangle$  &    
$\langle Age\rangle$ &        
$\langle$[Z/H]$\rangle$  &    
$\langle$\afe$\rangle$  &   
$S(\log(\rm Age))$ &    
$S($[Z/H]$)$ &       
$S([\alpha/{\rm Fe}])$    
\\
\tableline
$\z28-107$ & 1.935 & \z4.57 & 0.05 & 0.13 & $0.05-0.15$ & $0.07-0.12$ & $\leq0.13$ \\
$107-135$  & 2.088 & \z6.87 & 0.11 & 0.18 & $0.08-0.19$ & $0.06-0.10$ & $\leq0.10$ \\
$135-162$  & 2.173 & \z7.93 & 0.15 & 0.20 &  $\leq0.11$ & $0.07-0.11$ & $\leq0.08$ \\
$162-200$  & 2.257 & \z9.31 & 0.18 & 0.24 &  $\leq0.08$ & $0.06-0.10$ & $\leq0.09$ \\
$200-434$  & 2.386 &  10.82 & 0.24 & 0.28 & $0.03-0.08$ & $0.04-0.08$ & $\leq0.07$ \\
\tableline
\end{tabular}
\end{table*}

In the next section, we discuss our results from this section in
comparison with those of other studies, and consider the relation of
our results with recent observations of intermediate-redshift
clusters.

\section{Discussion}
\label{sec:discussion}

\subsection{Interpretation of the scaling relations}

In the following sections, we compare our results to other similar
studies and to observations at higher redshift. First, however, it is
useful to reiterate some caveats to interpreting the scaling relations
derived here.

As in any survey, the sample selection criteria must be borne in mind
when discussing our results. The NFPS sample analyzed here includes
only cluster members, based on the criteria of NFPS-I. Within each
cluster, the galaxies were selected by apparent magnitude (to $R=17$)
and color. Compared to some previous surveys, our sample probes to
fairly low mass, $\sigma\sim50$\,\kms.  The nonlinearity suggested by
Figure~\ref{fig:grinvert} is such that samples of more massive objects
would yield shallower age trends than studies covering the
low-$\sigma$ regime.  The color criterion rejects blue galaxies
further than 0.2\,mag from a red sequence fit to each cluster. While
such a cut should exclude actively star-forming galaxies, there is no
explicit selection on morphological type. Thus our sample includes
many S0 galaxies and some bulge-dominated spirals; such galaxies were
likely excluded (on a subjective basis) from many studies of
``bona-fide'' ellipticals.
 
To test for differences in the scaling relations between morphological
classes, we have used GIM2D \citep{Simard02} to derive a
bulge-to-total light ratios for about half of the galaxies in our
sample.  We see indications that ``diskier'' galaxies (i.e., those
with B/T$\la$0.5) follow a steeper age-$\sigma$ relation.  However,
there remains a significant age-$\sigma$ trend for the bulge-dominated
galaxies.  In a future paper, when final morphological information is
available for all of the galaxies, we will address this issue in much
greater detail.

An additional element in our selection process is the rejection, from
our {\sc CULL} sample, of galaxies with nebular emission lines. This
likely removes preferentially the later-type objects from our
analysis, but given that the emission selection depends in part on
\hb, it is important to test for any bias this introduces.  To
investigate the effects of emission selection, we have repeated the
analysis {\it including} galaxies with emission. In this case, we find
that the linestrength-$\sigma$ slopes for some of the metal lines
(e.g., Mg{\it b}) are on average steeper than the default solution and
thus our emission selection results in a shallower metallicity
gradient with $\sigma$.  However, the Balmer linestrength-$\sigma$
relations also steepen, suggesting that emission galaxies harbor young
underlying populations not totally disguised by emission in-filling.
As a result, including the emission galaxies yields an age trend even
stronger than in our default solution.  Other surveys {\it
corrected} their galaxies for emission \citep{Kunt01,Trager00b}
using the \oiiib\ correction factor of \cite{Trager00a} but still
included them in their analysis of the stellar population trends. We
have shown (Section~\ref{sec:emiss}) that such a scheme typically
under-corrects at low $\sigma$ and over-corrects at high $\sigma$.
The net effect of this will be to {\it flatten} the age-$\sigma$
relation derived in these studies.

The scaling relations for age, metallicity and \afe\ were determined
from central spectra. The fiber spectra sample a physical radius
0.2--0.8\,\hkpc, depending on the distance to each cluster, and
linestrength measurements were aperture-corrected to a common physical
radius of 0.68 \hkpc\ assuming universal gradients.  Although the
linestrength--$\sigma$ relations and the color--magnitude relation are
often considered as reflecting the same underlying trends, this
assumes, perhaps naively, that there is a trivial relationship between
{\it central} stellar populations and the {\it global} colors.

A final set of caveats concerns the stellar population models used to
translate the linestrength$-\sigma$ slopes into scaling relations of
population parameters.  For simplicity, these models describe the
highly idealized case of a single-age, single-metallicity
population. In elliptical galaxies, there may instead be a broad
distribution of metallicities \citep{Harris02}.  Similarly there could
be sub-populations of differing ages, as in the ``frosting'' models of
\cite{Trager00b}. Strictly, our analysis determines scaling relations
of the luminosity-weighted mean stellar age, metallicity and \afe.  An
alternative to a true age trend, therefore, would be a scenario in
which all galaxies have a 1\,Gyr ``frost'' representing a
progressively larger mass-fraction at lower velocity dispersion. A
separate concern is that the models, as described above, do not
include the effects of Blue Horizontal Branch (BHB) stars as observed
in metal-poor globular clusters. Note, however, that the evidence for
an age-trend is strongest in the case of \hb\ than for H$\delta_{\rm
F}$ and H$\gamma_{\rm F}$. Qualitatively, this is opposite to the
signature expected from variations in the BHB contribution
\citep{Schiavon04}. Robust constraints on this effect must await more
sophisticated models which simultaneously incorporate BHB stars and
$\alpha$-element abundances.

\subsection{Comparison to other results on scaling relations}

In this section, we compare our results with those from other groups,
taking into account the important differences in sample selection and
emission treatment as described above.

We have already noted that our results, particularly the age-$\sigma$
scaling relation, are not sensitive to which Balmer line (\hdf, \hgf,
\hb) is used.  This conclusion is reinforced in a separate paper,
\citep{Smith05}, which shows that a strong age gradient is also
required to reproduce the slope of the NFPS H$\alpha_{\rm A}-\sigma$
relation.

\cite{Trager00a} used principal component analysis to investigate
correlations of age, metallicity, and [E/Fe] (analogous to \afe) with
structural parameters in their sample of early-type galaxies in
clusters and in the field.  Although they model the scaling relations
in a different way, fitting metallicity and [E/Fe] to age and velocity
dispersion simultaneously, they find E/Fe $\propto \sigma^{0.33}$,
very similar to our value of \dafe$^{0.31}$.  Their metallicity
gradient is dependent on the individual ages of their galaxies, but
the velocity dispersion component is \dzh$^{0.76}$ compared to our
value of \dzh$^{0.53}$.  Although \cite{Trager00a} do not claim a
trend of age with $\sigma$, a simple fit to their data yields an
age-$\sigma$ trend with an exponent of $0.6\pm0.2$, consistent with
our results.

\cite{Kunt01}, using linestrengths for a sample drawn mostly from
clusters, explicitly assume no age trend with $\sigma$, which strongly
affects the other scaling relations derived. They find a higher \dzhd\ slope
($\sim$0.9) after correcting for varying \afe.  Their higher
metallicity slope and lack of an age slope is likely the effect of the
age-metallicity degeneracy, which we also see in the spread of points
in the leftmost panel of Figure~\ref{fig:ixefx}.  Modeling the
\cite{Kunt01} data according to our method, we 
obtain \dage$^{\approx 0.8}$ if all three parameters are allowed to
vary.  Although the relation is steeper than our default solution, 
it is quite sensitive to a single outlying low-$\sigma$ galaxy.  
Excluding this outlier (which has later-type morphology) yields 
\dage$^{\approx 0.5}$.

\cite{Poggianti01} obtained age and metallicity estimates of $\sim$280
red galaxies in the Coma Cluster.  
They found a broad range in age at all magnitudes, making it difficult 
to quantify the mean age-$\sigma$ relation, but they did note 
that the fraction of young dwarf galaxies in their sample
is higher than the fraction of young giant galaxies.  
Furthermore, in \cite{Poggianti04}, the authors noted that         
post-starburst k+a spectra were identified in dwarf galaxy spectra, 
with luminosities $L\lesssim0.1L*$. 
These results indicate a ``down-sizing'' effect in that the most
recent star formation activity occurs at lower redshifts for
progressively fainter galaxies.  This effect will be discussed further
in the next section.

\cite{Caldwell03} derived ages and metallicities for their sample of
175 early-type galaxies in clusters and in the field, including many
with $\sigma<100$\,\kms.  In Figure 21 of their paper, they plot age
versus $\sigma$, and note a strong correlation. Surprisingly, however,
these authors do not quote a numerical estimate for the slope.  From
their Table 9 and Figure 21, we estimate their log(age)-$\log \sigma$
gradient to be 0.8--1.2 depending on the index combination employed.
In addition, they find a shallower trend of metallicity than we do
(0.32 versus our value of 0.53).  Their age gradient is steeper than
ours, but their results lie on the age-metallicity degeneracy line for
these two parameters.

\cite{Thomas04} quote ages, metallicities and $\alpha$-enhancements
for 54 early-type galaxies in high-density environments.  They do not
quote scaling relations for the sample as a whole, but rather break
the data into subclasses pre-selected by age and velocity
dispersion. In order to compare their results with ours, we have
analyzed their published data for early-types in high-density
environments and derive the following scaling relations from a simple
unweighted regression on log $\sigma$: $0.78\pm0.23$, $0.42\pm0.14$
and $0.36\pm0.05$ for age, metallicity and $\alpha$-enhancement
respectively.  These results are in excellent agreement with our
results, but their errors are large because their sample contains few
low-$\sigma$ galaxies.

We note that, \cite{Proctor04} also found a positive age-$\sigma$
gradient in galaxies in Hickson Compact Groups.  They do not quote an
age-$\sigma$ relation but examination of their Figure 5 suggests a
relation somewhat steeper ($\sim 1.2$) than our best fit value.  

Figure~\ref{fig:compare} summarizes our results and those of
\cite{Trager00a},\cite{Caldwell03} and \cite{Thomas04}.  We conclude
that our scaling relations of age, metallicity and \afe\ fall within
the range spanned by previous studies. Differences among the above
results may arise in part from the different choices of linestrengths
used (resulting in scatter along the age--metallicity degeneracy
ellipse), and from different sample selection characteristics.

\subsection{Connection to observations at intermediate redshift}

The reality of our steep trend in age as a function of mass, along the
red-sequence, can be tested with observations at high redshift.  As
previously noted, the absolute age calibration of the models is less
secure than the relative ages. If we identify the stellar age of the
most massive systems ($\sigma\sim 400$ km s$^{-1}$; $\sim20\,L^*$) 
with 13 Gyr, close to the age of the Universe, then the least massive 
red-sequence galaxies, with $\sigma\approx60$\kms\ ($0.01\,L^*$) 
have ages of approximately 4\,Gyr, and galaxies with 
$\sigma\approx100$\kms\ ($\sim0.1\,L^*$) would 
have an ages $\sim$5.5\,Gyr.  Taking these at face value, and
considering also the substantial spread in age at given $\sigma$
(especially for low-$\sigma$ objects), we would expect to observe
strong evolutionary effects at intermediate redshifts.  At an earlier
epoch, some of the stellar mass presently residing on the faint end of
the red sequence will not have been formed at all; some will be
present in star-forming galaxies; and only a fraction will exist on
the red sequence itself. A key observational signature of such
evolution would be a depletion or truncation of the red sequence,
affecting progressively more massive galaxies with increasing lookback
time.

A number of studies have in fact advanced evidence for such a
depletion.  For example, \cite{Smail98} observed 10 clusters at
$z\approx0.24$, corresponding to a lookback time of 3.2 Gyr (for a
concordance cosmology).  Examination of their Figure 5 suggests a
strong decline in the number of red-sequence galaxies at $2.0-2.5$
magnitudes below $M^*$.  This corresponds to approximately
$\sigma\sim100$ km s$^{-1}$. Our age$-\sigma$ relation indicates {\it 
current} ages of 3--8\,Gyr for such objects, with a mean of 
$\sim5$\,Gyr.  Thus a substantial fraction of such galaxies either had 
not yet become quiescent at $z\sim0.24$, or else did not have enough
time to age onto the red sequence.  We conclude that the truncation of
the Smail et al. red sequence is at least approximately consistent
with the age$-\sigma$ relation obtained from NFPS.  At a somewhat
higher redshift of $z\approx0.75$ (lookback time of 7\,Gyr),
\cite{deLucia04} examined the red sequence in clusters and found a
deficiency of a factor of two (compared to Coma) for galaxies with
$0.1L^*<L<0.4L^*$.  Our age-$\sigma$ trend and scatter predicts that in
this luminosity range (corresponding to $\sigma\approx100-150$ km 
s$^{-1}$), the present-day stellar age is 7\,Gyr on average. Thus we would expect 
$\sim50\%$ of such galaxies to be significantly bluer at those 
epochs, again approximately consistent with the observed depletion.

For a $\sim$1\,Gyr period after star formation ceases, galaxies pass
through a post-starburst (or E+A) phase, characterized by very strong
Balmer absorption but no strong emission.  \cite{Tran03} studied E+A
post-starburst galaxies in three clusters at $z$=0.3, 0.6, and 0.8.
They find that the typical velocity dispersion of post-starburst
galaxies at these redshifts decreases from $\sim$170 km s$^{-1}$ at
$z$ = 0.8 to $\sim$100 km s$^{-1}$ at $z$ = 0.3.  If we identify these
objects as galaxies in the process of fading onto the red sequence,
then this trend of more massive galaxies becoming quiescent at higher
redshifts fits well with our present-day age-$\sigma$ trend.

In summary, our results add to recent evidence that the red sequence
of cluster galaxies has built up gradually over cosmic history,
progressing from more massive to less massive galaxies. Such a
scenario is the cluster analogue to the ``down-sizing'' of the
characteristic mass of star-forming field galaxies as discussed by
\cite{Cowie96} and \cite{Kauffmann04}.

\section{Conclusions}

In this paper, we have presented absorption linestrength measurements
for $\sim$5000 red-sequence galaxies in low-redshift clusters. Our
survey samples galaxies with velocity dispersion ranging down to
$\sim$50\,\kms.  The absorption data are complemented with emission
line measurements which can be used to select subsamples with only low
levels of nebular contamination.  We have employed the slopes of the
linestrength--velocity dispersion relations to constrain linear
scaling relations for stellar population parameters. We find that more
massive galaxies are older, have higher overall metallicity, and have
higher \afe\ ratios than galaxies of lower mass (strictly, this refers
to central, luminosity-averaged properties of the stellar
populations). These conclusions are quite robust, and in particular
are not dependent on which of the age-sensitive Balmer lines are used
in the analysis. Moreover, a more traditional model-grid inversion
yields consistent scaling relations.

The most important result of this paper is the very clear detection of
an apparent age--mass relation for red sequence galaxies in
clusters. This arises because a flat age-$\sigma$ relation cannot
generate the steep slopes of the Balmer linestrength-$\sigma$
relations while simultaneously matching the metal
linestrength-$\sigma$ slopes.  This is consistent with the results
from a number of other independent studies based on smaller samples.

A strong age--mass relation stands in stark contrast to the widespread
assumption that cluster ellipticals form a metallicity sequence of
approximately constant age.  Moreover, our apparently
anti-hierarchical age--mass relation disagrees with the predictions
from semi-analytic galaxy formation models, which suggest either that 
brighter ellipticals have slightly younger stellar populations \citep{Kauffmann98} 
or that early-type cluster galaxies should have uniformly old stellar
populations (Fig. 18 of \cite{Kunt02}, which is based upon the models
of \cite{Cole00}).  

Our results are broadly consistent with claims of a truncated or
depleted red sequence in clusters at higher redshifts (e.g., Smail et
al. 1998; de Lucia et al. 2004), and of an increase in the average
masses of post-starburst cluster members with increasing redshift
\citep{Tran03}. Together, these observations suggest a trend of
``down-sizing'' galaxy formation in clusters, mirroring a similar
decline in the characteristic mass of star-forming field galaxies
\citep{Cowie96,Kauffmann04}.  These studies all present a picture of
the red sequence in clusters building up slowly over cosmic history,
proceeding from the most massive to progressively lower mass
galaxies. In particular, galaxies presently on the faint end of the
red sequence became quiescent only at very recent epochs, and are
likely the descendants of star-forming or post-starburst galaxies in
intermediate-redshift clusters.

\section*{ACKNOWLEDGMENTS}
We gratefully acknowledge the substantial assignment of NOAO observing
resources to the NFPS program and a Chretien International Research
Grant from the American Astronomical Society.  JEN was supported by a
Dartmouth College Fellowship, a Graduate Assistance in Areas of
National Need fellowship, and a NASA space grant.  MJH acknowledges
support from the NSERC of Canada, an Ontario Premier's Research
Excellence Award. SAWM was supported from the PPARC grant
``Extragalactic Astronomy \& Cosmology at Durham 1998-2002.''  SJQ was
supported by a PPARC studentship.  IRAF is distributed by the National
Optical Astronomy Observatories which is operated by the Association
of Universities for Research in Astronomy, Inc. under contract with
the National Science Foundation.

\bibliographystyle{apj}
\bibliography{ms}

\begin{thebibliography}{60}
\expandafter\ifx\csname natexlab\endcsname\relax\def\natexlab#1{#1}\fi

\bibitem[{{Abazajian} {et~al}(2004)}]{SDSSdr2}
{Abazajian}, K. {et~al} 2004, \aj, 128, 502

\bibitem[{{Balogh} {et~al.}(1999){Balogh}, {Morris}, {Yee}, {Carlberg}, \&
  {Ellingson}}]{Balogh99}
{Balogh}, M.~L., {Morris}, S.~L., {Yee}, H.~K.~C., {Carlberg}, R.~G., \&
  {Ellingson}, E. 1999, \apj, 527, 54

\bibitem[{{Bender} {et~al.}(1993){Bender}, {Burstein}, \& {Faber}}]{Bender93}
{Bender}, R., {Burstein}, D., \& {Faber}, S.~M. 1993, \apj, 411, 153

\bibitem[{{Bernardi} {et~al}(2003)}]{Bernardi03}
{Bernardi}, M. {et~al} 2003, \aj, 125, 1882

\bibitem[{{Bower} {et~al.}(1992){Bower}, {Lucey}, \& {Ellis}}]{Bower92}
{Bower}, R.~G., {Lucey}, J.~R., \& {Ellis}, R.~S. 1992, \mnras, 254, 601

\bibitem[{{Burstein} {et~al.}(1984){Burstein}, {Faber}, {Gaskell}, \&
  {Krumm}}]{Burstein84}
{Burstein}, D., {Faber}, S.~M., {Gaskell}, C.~M., \& {Krumm}, N. 1984, \apj,
  287, 586

\bibitem[{{Caldwell} {et~al.}(2003){Caldwell}, {Rose}, \&
  {Concannon}}]{Caldwell03}
{Caldwell}, N., {Rose}, J.~A., \& {Concannon}, K.~D. 2003, \aj, 125, 2891

\bibitem[{{Cardiel} {et~al.}(1998){Cardiel}, {Gorgas}, {Cenarro}, \&
  {Gonzalez}}]{Cardiel98}
{Cardiel}, N., {Gorgas}, J., {Cenarro}, J., \& {Gonzalez}, J.~J. 1998, \aaps,
  127, 597

\bibitem[{{Cole} {et~al.}(2000){Cole}, {Lacey}, {Baugh}, \& {Frenk}}]{Cole00}
{Cole}, S., {Lacey}, C.~G., {Baugh}, C.~M., \& {Frenk}, C.~S. 2000, \mnras,
  319, 168

\bibitem[{{Colless} {et~al.}(1999){Colless}, {Burstein}, {Davies}, {McMahan},
  {Saglia}, \& {Wegner}}]{Colless99}
{Colless}, M., {Burstein}, D., {Davies}, R.~L., {McMahan}, R.~K., {Saglia},
  R.~P., \& {Wegner}, G. 1999, \mnras, 303, 813

\bibitem[{{Cowie} {et~al.}(1996){Cowie}, {Songaila}, {Hu}, \&
  {Cohen}}]{Cowie96}
{Cowie}, L.~L., {Songaila}, A., {Hu}, E.~M., \& {Cohen}, J.~G. 1996, \aj, 112,
  839

\bibitem[{{Davies} {et~al.}(1987){Davies}, {Burstein}, {Dressler}, {Faber},
  {Lynden-Bell}, {Terlevich}, \& {Wegner}}]{7S}
{Davies}, R.~L., {Burstein}, D., {Dressler}, A., {Faber}, S.~M., {Lynden-Bell},
  D., {Terlevich}, R.~J., \& {Wegner}, G. 1987, \apjs, 64, 581

\bibitem[{{De Lucia} {et~al}(2004)}]{deLucia04}
{De Lucia}, G. {et~al} 2004, \apjl, 610, L77

\bibitem[{{Djorgovski} \& {Davis}(1987)}]{Djorgovski87}
{Djorgovski}, S., \& {Davis}, M. 1987, \apj, 313, 59

\bibitem[{{Dressler}(1987)}]{Dressler87}
{Dressler}, A. 1987, \apj, 317, 1

\bibitem[{{Ebeling} {et~al.}(1998){Ebeling}, {Edge}, {Bohringer}, {Allen},
  {Crawford}, {Fabian}, {Voges}, \& {Huchra}}]{Ebeling98}
{Ebeling}, H., {Edge}, A.~C., {Bohringer}, H., {Allen}, S.~W., {Crawford},
  C.~S., {Fabian}, A.~C., {Voges}, W., \& {Huchra}, J.~P. 1998, \mnras, 301,
  881

\bibitem[{{Ebeling} {et~al.}(1996){Ebeling}, {Voges}, {Bohringer}, {Edge},
  {Huchra}, \& {Briel}}]{Ebeling96}
{Ebeling}, H., {Voges}, W., {Bohringer}, H., {Edge}, A.~C., {Huchra}, J.~P., \&
  {Briel}, U.~G. 1996, \mnras, 281, 799

\bibitem[{{Faber} \& {Jackson}(1976)}]{faber76}
{Faber}, S.~M., \& {Jackson}, R.~E. 1976, \apj, 204, 668

\bibitem[{{Gonz{\' a}lez}(1993)}]{Gonzalez93}
{Gonz{\' a}lez}, J.~J. 1993, PhD thesis, University of California, Santa Cruz

\bibitem[{{Goudfrooij} {et~al.}(1994){Goudfrooij}, {de Jong}, {Hansen}, \&
  {Norgaard-Nielsen}}]{Goudfrooij94}
{Goudfrooij}, P., {de Jong}, T., {Hansen}, L., \& {Norgaard-Nielsen}, H.~U.
  1994, \mnras, 271, 833

\bibitem[{{Harris} \& {Harris}(2002)}]{Harris02}
{Harris}, W.~E., \& {Harris}, G.~L.~H. 2002, \aj, 123, 3108

\bibitem[{{Herbig} \& {Mayall}(1957)}]{Mayall57}
{Herbig}, G.~H., \& {Mayall}, N.~U. 1957, \pasp, 69, 563

\bibitem[{{Jorgensen}(1997)}]{Jorg97}
{Jorgensen}, I. 1997, \mnras, 288, 161

\bibitem[{{Jorgensen} {et~al.}(1995){Jorgensen}, {Franx}, \&
  {Kjaergaard}}]{Jorg95}
{Jorgensen}, I., {Franx}, M., \& {Kjaergaard}, P. 1995, \mnras, 276, 1341

\bibitem[{{Kauffmann} \& {Charlot}(1998)}]{Kauffmann98}
{Kauffmann}, G., \& {Charlot}, S. 1998, \mnras, 294, 705

\bibitem[{{Kauffmann} {et~al.}(2004){Kauffmann}, {White}, {Heckman}, {M{\'
  e}nard}, {Brinchmann}, {Charlot}, {Tremonti}, \& {Brinkmann}}]{Kauffmann04}
{Kauffmann}, G., {White}, S.~D.~M., {Heckman}, T.~M., {M{\' e}nard}, B.,
  {Brinchmann}, J., {Charlot}, S., {Tremonti}, C., \& {Brinkmann}, J. 2004,
  \mnras, 353, 713

\bibitem[{{Kinney} {et~al.}(1996){Kinney}, {Calzetti}, {Bohlin}, {McQuade},
  {Storchi-Bergmann}, \& {Schmitt}}]{Kinney96}
{Kinney}, A.~L., {Calzetti}, D., {Bohlin}, R.~C., {McQuade}, K.,
  {Storchi-Bergmann}, T., \& {Schmitt}, H.~R. 1996, \apj, 467, 38

\bibitem[{{Kuntschner} {et~al.}(2001){Kuntschner}, {Lucey}, {Smith}, {Hudson},
  \& {Davies}}]{Kunt01}
{Kuntschner}, H., {Lucey}, J.~R., {Smith}, R.~J., {Hudson}, M.~J., \& {Davies},
  R.~L. 2001, \mnras, 323, 615

\bibitem[{{Kuntschner} {et~al.}(2002){Kuntschner}, {Smith}, {Colless},
  {Davies}, {Kaldare}, \& {Vazdekis}}]{Kunt02}
{Kuntschner}, H., {Smith}, R.~J., {Colless}, M., {Davies}, R.~L., {Kaldare},
  R., \& {Vazdekis}, A. 2002, \mnras, 337, 172

\bibitem[{{Maraston}(1998)}]{Maraston98}
{Maraston}, C. 1998, \mnras, 300, 872

\bibitem[{{Maraston} \& {Thomas}(2000)}]{Maraston00}
{Maraston}, C., \& {Thomas}, D. 2000, \apj, 541, 126

\bibitem[{{Moore} {et~al.}(2002){Moore}, {Lucey}, {Kuntschner}, \&
  {Colless}}]{Moore02}
{Moore}, S.~A.~W., {Lucey}, J.~R., {Kuntschner}, H., \& {Colless}, M. 2002,
  \mnras, 336, 382

\bibitem[{{Phillips} {et~al.}(1986){Phillips}, {Jenkins}, {Dopita}, {Sadler},
  \& {Binette}}]{Phillips86}
{Phillips}, M.~M., {Jenkins}, C.~R., {Dopita}, M.~A., {Sadler}, E.~M., \&
  {Binette}, L. 1986, \aj, 91, 1062

\bibitem[{{Poggianti} {et~al.}(2004){Poggianti}, {Bridges}, {Komiyama}, {Yagi},
  {Carter}, {Mobasher}, {Okamura}, \& {Kashikawa}}]{Poggianti04}
{Poggianti}, B.~M., {Bridges}, T.~J., {Komiyama}, Y., {Yagi}, M., {Carter}, D.,
  {Mobasher}, B., {Okamura}, S., \& {Kashikawa}, N. 2004, \apj, 601, 197

\bibitem[{{Poggianti} {et~al.}(2001){Poggianti}, {Bridges}, {Mobasher},
  {Carter}, {Doi}, {Iye}, {Kashikawa}, {Komiyama}, {Okamura}, {Sekiguchi},
  {Shimasaku}, {Yagi}, \& {Yasuda}}]{Poggianti01}
{Poggianti}, B.~M., {Bridges}, T.~J., {Mobasher}, B., {Carter}, D., {Doi}, M.,
  {Iye}, M., {Kashikawa}, N., {Komiyama}, Y., {Okamura}, S., {Sekiguchi}, M.,
  {Shimasaku}, K., {Yagi}, M., \& {Yasuda}, N. 2001, \apj, 562, 689

\bibitem[{{Proctor}(2002)}]{Proctor02}
{Proctor}, R.~N. 2002, PhD thesis, University of Central Lancashire

\bibitem[{{Proctor} {et~al.}(2004{\natexlab{a}}){Proctor}, {Forbes}, \&
  {Beasley}}]{Proctorchi2}
{Proctor}, R.~N., {Forbes}, D.~A., \& {Beasley}, M.~A. 2004{\natexlab{a}},
  \mnras, 355, 1327

\bibitem[{{Proctor} {et~al.}(2004{\natexlab{b}}){Proctor}, {Forbes}, {Hau},
  {Beasley}, {De Silva}, {Contreras}, \& {Terlevich}}]{Proctor04}
{Proctor}, R.~N., {Forbes}, D.~A., {Hau}, G.~K.~T., {Beasley}, M.~A., {De
  Silva}, G.~M., {Contreras}, R., \& {Terlevich}, A.~I. 2004{\natexlab{b}},
  \mnras, 349, 1381

\bibitem[{{Saglia} {et~al.}(2002){Saglia}, {Maraston}, {Thomas}, {Bender}, \&
  {Colless}}]{Saglia02}
{Saglia}, R.~P., {Maraston}, C., {Thomas}, D., {Bender}, R., \& {Colless}, M.
  2002, \apjl, 579, L13

\bibitem[{{Sandage} \& {Visvanathan}(1978)}]{Sandage78}
{Sandage}, A., \& {Visvanathan}, N. 1978, \apj, 225, 742

\bibitem[{{Schiavon} {et~al.}(2004){Schiavon}, {Rose}, {Courteau}, \&
  {MacArthur}}]{Schiavon04}
{Schiavon}, R.~P., {Rose}, J.~A., {Courteau}, S., \& {MacArthur}, L.~A. 2004,
  \apjl, 608, L33

\bibitem[{{Simard} {et~al.}(2002){Simard}, {Willmer}, {Vogt}, {Sarajedini},
  {Phillips}, {Weiner}, {Koo}, {Im}, {Illingworth}, \& {Faber}}]{Simard02}
{Simard}, L., {Willmer}, C.~N.~A., {Vogt}, N.~P., {Sarajedini}, V.~L.,
  {Phillips}, A.~C., {Weiner}, B.~J., {Koo}, D.~C., {Im}, M., {Illingworth},
  G.~D., \& {Faber}, S.~M. 2002, \apjs, 142, 1

\bibitem[{{Smail} {et~al.}(1998){Smail}, {Edge}, {Ellis}, \&
  {Blandford}}]{Smail98}
{Smail}, I., {Edge}, A.~C., {Ellis}, R.~S., \& {Blandford}, R.~D. 1998, \mnras,
  293, 124

\bibitem[{{Smith}(2005)}]{Smith05}
{Smith}, R.~J. 2005, {MNRAS, submitted}

\bibitem[{{Smith} {et~al.}(2004){Smith}, {Hudson}, {Nelan}, {Moore}, {Quinney},
  {Wegner}, {Lucey}, {Davies}, {Malecki}, {Schade}, \& {Suntzeff}}]{Smith04}
{Smith}, R.~J., {Hudson}, M.~J., {Nelan}, J.~E., {Moore}, S.~A.~W., {Quinney},
  S.~J., {Wegner}, G.~A., {Lucey}, J.~R., {Davies}, R.~L., {Malecki}, J.~J.,
  {Schade}, D., \& {Suntzeff}, N.~B. 2004, \aj, 128, 1558

\bibitem[{{Stasi{\' n}ska} {et~al.}(2004){Stasi{\' n}ska}, {Mateus}, {Sodr{\'
  e}}, \& {Szczerba}}]{Stasi04}
{Stasi{\' n}ska}, G., {Mateus}, A., {Sodr{\' e}}, L., \& {Szczerba}, R. 2004,
  \aap, 420, 475

\bibitem[{{Terlevich} \& {Forbes}(2002)}]{Terlevich02}
{Terlevich}, A.~I., \& {Forbes}, D.~A. 2002, \mnras, 330, 547

\bibitem[{{Thomas} {et~al.}(2003){Thomas}, {Maraston}, \& {Bender}}]{TMB03}
{Thomas}, D., {Maraston}, C., \& {Bender}, R. 2003, \mnras, 339, 897

\bibitem[{{Thomas} {et~al.}(2004{\natexlab{a}}){Thomas}, {Maraston}, {Bender},
  \& {Mendes de Oliveira}}]{Thomas04}
{Thomas}, D., {Maraston}, C., {Bender}, R., \& {Mendes de Oliveira}, C.
  2004{\natexlab{a}}, {ApJ, submitted}, {astro-ph/0410209}

\bibitem[{{Thomas} {et~al.}(2004{\natexlab{b}}){Thomas}, {Maraston}, \&
  {Korn}}]{TMK04}
{Thomas}, D., {Maraston}, C., \& {Korn}, A. 2004{\natexlab{b}}, \mnras, 351,
  L19

\bibitem[{{Trager} {et~al.}(2000{\natexlab{a}}){Trager}, {Faber}, {Worthey}, \&
  {Gonz{\' a}lez}}]{Trager00b}
{Trager}, S.~C., {Faber}, S.~M., {Worthey}, G., \& {Gonz{\' a}lez}, J.~J.
  2000{\natexlab{a}}, \aj, 120, 165

\bibitem[{{Trager} {et~al.}(2000{\natexlab{b}}){Trager}, {Faber}, {Worthey}, \&
  {Gonz{\' a}lez}}]{Trager00a}
---. 2000{\natexlab{b}}, \aj, 119, 1645

\bibitem[{{Trager} {et~al.}(1998){Trager}, {Worthey}, {Faber}, {Burstein}, \&
  {Gonzalez}}]{Trager98}
{Trager}, S.~C., {Worthey}, G., {Faber}, S.~M., {Burstein}, D., \& {Gonzalez},
  J.~J. 1998, \apjs, 116, 1

\bibitem[{{Tran} {et~al.}(2003){Tran}, {Franx}, {Illingworth}, {Kelson}, \&
  {van Dokkum}}]{Tran03}
{Tran}, K.~H., {Franx}, M., {Illingworth}, G., {Kelson}, D.~D., \& {van
  Dokkum}, P. 2003, \apj, 599, 865

\bibitem[{{van Dokkum} \& {Ellis}(2003)}]{vanDokkum03}
{van Dokkum}, P.~G., \& {Ellis}, R.~S. 2003, \apjl, 592, L53

\bibitem[{{Vazdekis}(1999)}]{Vazdekis99}
{Vazdekis}, A. 1999, \apj, 513, 224

\bibitem[{{Wegner} {et~al.}(1999){Wegner}, {Colless}, {Saglia}, {McMahan},
  {Davies}, {Burstein}, \& {Baggley}}]{Wegner99}
{Wegner}, G., {Colless}, M., {Saglia}, R.~P., {McMahan}, R.~K., {Davies},
  R.~L., {Burstein}, D., \& {Baggley}, G. 1999, \mnras, 305, 259

\bibitem[{{Worthey}(1994)}]{Worthey94}
{Worthey}, G. 1994, \apjs, 95, 107

\bibitem[{{Worthey} \& {Ottaviani}(1997)}]{Worthey97}
{Worthey}, G., \& {Ottaviani}, D.~L. 1997, \apjs, 111, 377

\bibitem[{{Wuyts} {et~al.}(2004){Wuyts}, {van Dokkum}, {Kelson}, {Franx}, \&
  {Illingworth}}]{Wuyts04}
{Wuyts}, S., {van Dokkum}, P.~G., {Kelson}, D.~D., {Franx}, M., \&
  {Illingworth}, G.~D. 2004, \apj, 605, 677

\end{thebibliography}

\label{lastpage}

\end{document}